\documentclass{aa}  
\usepackage{xcolor}
\usepackage{graphicx}
\usepackage{txfonts}
\begin{document}

   \title{Photoprocessing of H$_2$S on dust grains: building S chains in translucent clouds and comets}


   \author{S. Cazaux\inst{1,2}
          \and
          H. Carrascosa\inst{3}
          \and
          G. M. Mu\~noz Caro\inst{3}
          \and
          P. Caselli\inst{4} 
          \and
          A. Fuente\inst{5}
          \and
          D. Navarro-Almaida\inst{5}
          \and
          P. Riviére-Marichalar\inst{5}
          }

   \institute{Faculty of Aerospace Engineering, Delft University of Technology, Delft, The Netherlands
        \email{s.m.cazaux@tudelft.nl}\\
        \and
        Leiden Observatory, Leiden University, P.O. Box 9513, NL 2300 RA Leiden, The Netherlands\\
        \and
         Centro de Astrobiolog\'{\i}a (CSIC-INTA), Ctra. de Ajalvir, km 4, Torrej\'on de Ardoz, 28850 Madrid, Spain\\
         \and
         Max Planck Institute for Extraterrestrial Physics, Postfach 1312, 85741 Garching, Germany\\
         \and
         Observatorio Astron\'omico Nacional (OAN, IGN), Apdo 112, 28803 Alcal\'a de Henares, Spain}

   \date{Received September 15, 1996; accepted March 16, 1997}

 
  \abstract
   {Sulfur is a biogenic element used as a tracer of the evolution from interstellar clouds to stellar systems. However, most of the expected sulfur in molecular clouds remains undetected. Sulfur disappears from the gas phase in two steps. One first depletion occurs during the translucent phase, reducing the gas phase sulfur by 7-40 times, while the following freeze-out step occurs in molecular clouds, reducing it by another order of magnitude. This long-standing dilemma awaits an explanation.}
   {The aim of this study is to understand under which form the missing sulfur is hiding in molecular clouds. The possibility that sulfur is depleted onto dust grains is considered.}
   {Experimental simulations mimicking H$_2$S ice UV-photoprocessing in molecular clouds were conducted at 8 K under ultra-high vacuum. The ice was subsequently warmed up to room temperature. The ice was monitored using infrared spectroscopy and the desorbing molecules were measured by quadrupole mass spectrometry in the gas phase. Theoretical Monte Carlo simulations were performed for interpretation of the experimental results and extrapolation to the astrophysical and planetary conditions.}
   {H$_2$S$_2$ formation was observed during irradiation at 8 K. Molecules H$_2$S$_x$ with $x$ > 2 were also identified and found to desorb during warm-up, along with S$_2$ to S$_4$ species. Larger S$_x$ molecules up to S$_8$ are refractory at room temperature and remained on the substrate forming a residue. Monte Carlo simulations were able to reproduce the molecules desorbing during warming up, and found that residues are chains or sulfur consisting of 6-7 atoms.}
   {Based on the interpretation of the experimental results using our theoretical model, it is proposed that S$^+$ in translucent clouds contributes notoriously to S depletion in denser regions by forming long S-chains on dust grains in few times 10$^4$ years. We suggest that the S$_2$ to S$_4$ molecules observed in comets are not produced by fragmentation of these large chains. Instead, they probably come  either from UV-photoprocessing of H$_2$S-bearing ice produced in molecular clouds or from short S chains formed during the translucent cloud phase.}


   \maketitle
%

\section{Introduction}
Sulfur is an important element for life as we know it and its chemistry is particularly relevant for linking interstellar clouds to star forming regions, to protoplanetary disks and stellar systems such as our own. S-bearing molecules are frequently used to trace the kinematics and the chemical evolution of star- \citep[e.g.][]{Zhou1993,Wakelam2004} and planet-forming regions \citep[e.g.][]{Dutrey1997,LeGal2019}. Several S-bearing species have also been recently detected in the coma of comet 67P \citep{calmonte2016}. There is however a big problem related to sulfur chemistry, which was first recognized in the early 70s: it is not yet clear in which form most of the sulfur resides in molecular clouds.  In fact, while cosmic abundances of (ionized) sulfur are found in diffuse clouds \citep{Jenkins2009}, the abundance of the most abundant S-bearing molecule in molecular clouds, CS, cannot be reproduced by astrochemical models, unless the S abundance in the gas phase is reduced by several orders of magnitude \citep[e.g.][]{Penzias1971,Oppenheimer1974,hasegawa1992,Bulut2021}. 

Several attempts have been done in the past to shed light on the "missing sulfur" problem in molecular clouds. \citet{Caselli1994} considered the possibility that most of the S in molecular clouds is residing on the surface of dust grains; if this is the case, the rapid formation of H$_2$S, and the production of H$_2$ from the H$_2$S+H reaction in the ice, \rm{create a sink of H atoms,}\rm\ which could also explain the very slow conversion of CO ice in CH$_3$OH, which is indeed only observed in dense cores of molecular clouds and star-forming regions \citep[e.g.][]{Boogert2015,Goto2020}. However, only upper limits have been measured for H$_2$S in ice \citep[e.g.][]{Smith1991,vanderTak2003,jimenez2011}, thus excluding that the majority of solid S is in H$_2$S form. \citet{Ruffle1999} proposed that most of the S depletion happens in translucent clouds, where S atoms are still mainly in atomic ionized form \citep[see also][]{Sternberg1995} while the dust grains are mainly negatively charged; in this scenario, the Coulomb-enhanced freeze-out rate of S$^+$ will rapidly deplete sulfur from the gas phase \citep[see also][for generic cations]{Umebayashi1980}. \citet{Vidal2017} compared their astrochemical modeling results to observations toward the TMC-1(CP) dark cloud and concluded that, depending on the cloud age, most of the sulfur could be either in (the not observable) atomic form or on HS and H$_2$S in solid form. \citet{laas2019} expanded the chemistry of sulfur by including S-bearing organic molecules and the Ruffle et al. (1999) freeze-out mechanism; they found a good agreement with observations in molecular clouds and concluded that the majority of S is trapped in organo-sulfur species in icy mantles of dust grains. 

In laboratory experiments where H$_2$S ice is irradiated either with UV photons or energetic particles (proxy for the interstellar cosmic rays), it has been found that a significant fraction of H$_2$S can be transformed in allotropic forms of S, including the most stable form S$_8$ \citep[e.g.][]{jimenez2011,jimenez2014}. This has also been confirmed theoretically by \citet{Shingledecker2020} using astrochemical models inclusive of radiation chemistry; interestingly, these models predict that the impact of energetic particles in H$_2$S-rich ices reduces the abundance of H$_2$S, while enhancing SO$_2$, OCS and S$_8$. The first interstellar molecule containing more than one S atom (S$_2$H) was detected recently by \citet{Fuente2017} toward the Horsehead photodissociation region, suggesting an interesting interplay between gas-phase and surface chemistry. The GEMS large project \citep{RBaras2021} at the IRAM 30m telescope is also providing important observational constraints on sulfur chemistry in molecular clouds; for example \citet{navarro2020} measured gas-phase H$_2$S across several sources, finding good agreement with chemical models assuming non depleted S abundances, which however highly overestimate the CS abundance. With the ROSINA instrument onboard Rosetta, \citet{calmonte2016} measured a variety of S-bearing molecules in comet 67P, which appears to preserve pre-stellar ice \citep{Altwegg2019,Drozdovskaya2019,Drozdovskaya2021}: the majority of S is in H$_2$S ice (about 57\%), followed by atomic S, SO$_2$, SO, OCS, H$_2$CS, CH$_3$SH, CS$_2$, S$_2$ and other S-bearing organics. 

Despite the many observations and chemical models available, there are still many open questions about sulfur chemistry in the interstellar medium: what happens to S atoms when adsorbed onto dust grains in translucent and molecular clouds? What is the chemistry of irradiated H$_2$S ice? Which fraction of sulfur atoms will form chains? Which S-bearing molecules should be observed to test the models? In this paper we focus our attention on detailed experimental work and its simulation with a Monte Carlo chemical model, with the aim of putting quantitative constraints on the S depletion and chemical processing in translucent and molecular clouds. 

The paper is structured as follows:  Section\,\ref{Sect:experiments} describes the laboratory experiments and  Section\,\ref{Sect:exp_results} presents the results of the experimental work; Monte Carlo simulations are described in Section\,\ref{Sect:MonteCarlo}. Section\,\ref{Sect:photo} discusses photo-processes of H$_2$S ice in simulations and experimental work, while applications to interstellar molecular clouds and to Solar System conditions can be found in Sections\,\ref{trans} and \ref{PS}, respectively. Our conclusions are in Section\,\ref{Sect:conclusions}.

\section{Experiments: H$_2$S irradiation with UV photons} \label{Sect:experiments}
The experimental results presented here have been obtained using the InterStellar Astrochemistry Chamber (ISAC) at the Centro de
Astrobiología \citep{munozcaro2010}. ISAC is an ultra-high
vacuum (UHV) chamber with a base pressure dominated by background H$_2$ in the range $P$ = 2.5-4.0 x 10$^{-11}$ mbar, which corresponds to dense cloud interiors.
H$_2$S ice samples were grown by accretion of gas molecules onto the tip of a cold finger at 8 K, where a MgF$_2$ infrared (IR) and ultraviolet (UV) transparent window serves as the substrate for deposition. This temperature was achieved by means of a closed-cycle helium cryostat. The samples were then warmed up until a complete sublimation was attained. A silicon diode temperature sensor, and a LakeShore Model 331 temperature controller are used, reaching an accuracy of about 0.1 K. 

The evolution of the solid sample was monitored by in situ transmittance Fourier Transform Infrared spectroscopy (FTIR) with a spectral resolution of 2 cm$^{-1}$. Column densities of each species in the ice were calculated from the IR spectra using the formula\\
\begin{equation}
 N = \frac{1}{A} \int_{band} {\tau_{\nu}d\nu},
\label{N}
\end{equation}
where $N$ is the column density in cm$^{-2}$, ${\tau}_{\nu}$ the optical depth of the band, $d\nu$ the wavenumber differential in cm$^{-1}$, and $A$ the band strength in cm molecule$^{-1}$. The integrated absorbance is 
equal to 0.43 $\times$ $\tau$, where $\tau$ is the integrated optical 
depth of the band. The adopted band strengths, $A$, derived from laboratory experiments are provided in Table 1. 

\begin{table}[htbp]
    \begin{center}
    \begin{tabular}{|c|c|c|c|c|}
    \hline
    Species &  Feature  &Position& Position & $A$$\times$ 10$^{-17}$ \\
            &           &   $\mu$m  &   cm$^{-1}$ & cm molec$^{-1}$  \\
    \hline \hline
    H$_2$S	  & S-H stretch	 &    3.940     & 2538  &   2.0	 $^a$ \\ \hline 
    H$_2$S$_2$  & S-H stretch	 &    4.016    & 2490  &  2.4$\pm$0.5		$^b$ \\ 
    \hline 
\end{tabular}
\end{center}
$^a$ \cite{jimenez2011}\\  
$^b$ This work. Band strength of H$_2$S$_2$ was calculated between 100 K and 120 K, showing a constant value in this range, see Sect. \ref{sect_warm-up}.\\ 
\caption{Photo-dissociation reactions with associated rates considered in the present study. These rates are taken from \cite{Shingledecker2020}. }
\label{diss}
\end{table}


The accretion rate measured by IR spectroscopy of the deposited ice was 1 ML s$^{-1}$. 
Samples were UV-irradiated using a microwave-stimulated hydrogen flow discharge lamp (MDHL) that provides a flux of 2.5 $\times$ 10$^{14}$ photons cm$^{-2}$ s$^{-1}$ at 
the sample position with an average photon energy of 8.6 eV. The MDHL spectrum is reported in Cruz-Diaz et al. (2014). UV spectra were measured routinely {\em in situ} during the experiments with the use of a McPherson 0.2 meter focal length VUV monochromator (Model 234/302) with a photomultiplier tube (PMT) detector equipped with a sodium salicylate window, optimized to operate from 100-500 nm (11.27-2.47 eV), with a spectral resolution of 0.4 nm. UV absorption spectra of H$_2$S ice samples served to monitor the detection of photoproducts generated upon UV-irradiation at 8 K or later during warm-up. For more details on the experimental protocol employed for VUV spectroscopy, see Cruz-Diaz et al. (2014).

\section{Experimental results} \label{Sect:exp_results}

\begin{figure}
  \centering 
  \includegraphics[width=0.5\textwidth]{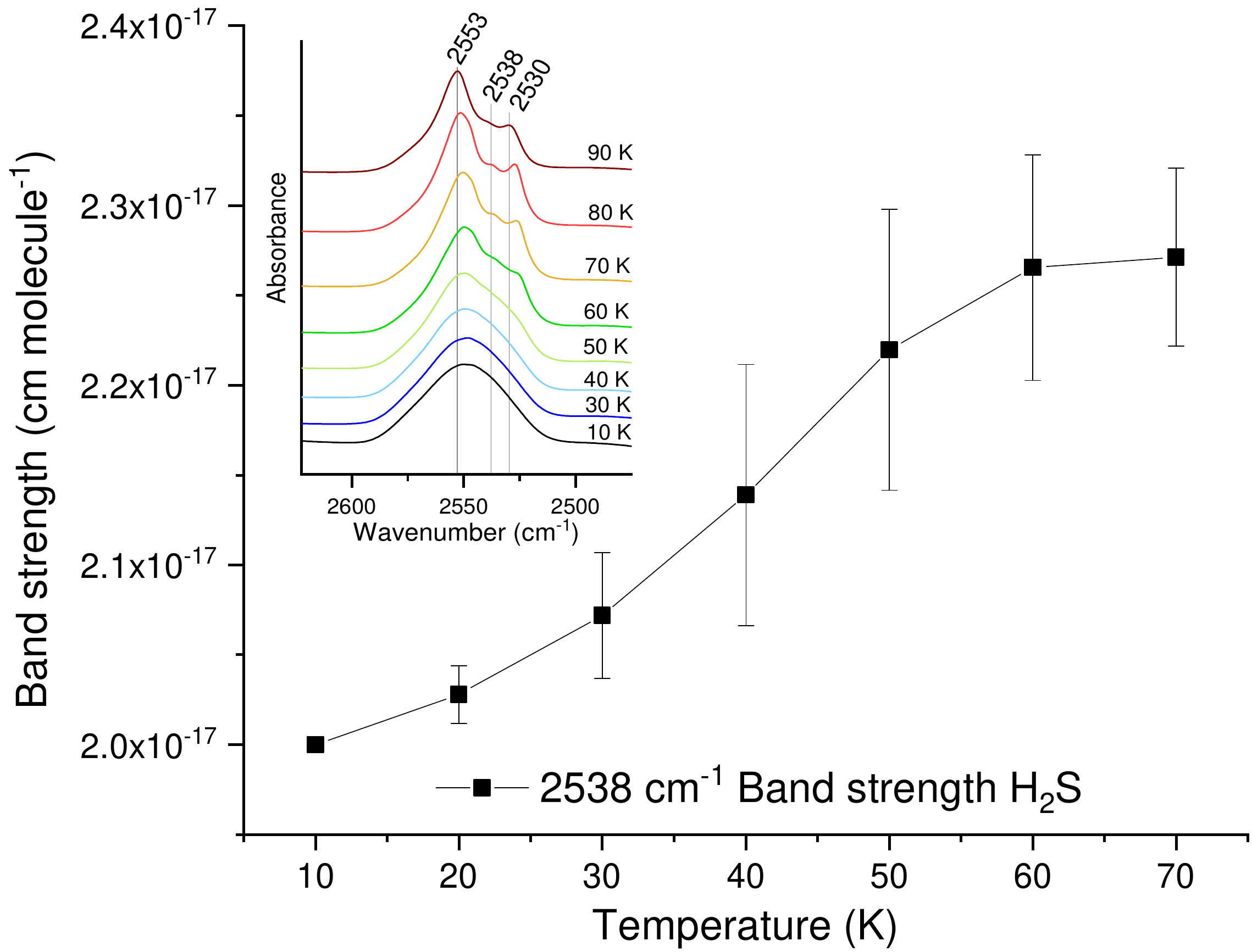}
  \caption{Evolution of the H$_2$S band strength as a function of temperature. Above 70 K, thermal desorption becomes important, reducing H$_2$S column density, preventing us from a quantification of the band strength.}
  \label{Fig.IR_band_strength_H2S}
\end{figure}

As reported by \cite{jimenez2011}, thermal processing of H$_2$S ice leads to crystallization at temperatures above 40 K. This process reduces the number of randomly oriented molecules and the vibration modes become narrower, see inlet of Fig. \ref{Fig.IR_band_strength_H2S}, which agrees with Fig. 1 in \cite{jimenez2011}. The antisymmetric and symmetric vibrations appear at 2553 and 2530 cm$^{-1}$, respectively. The remaining amorphous domains are responsible for the IR band centered at 2538 cm$^{-1}$.

\subsection{Photoproduct formation in UV-irradiated H$_2$S ice}
\label{sect.UV-irradiation_H2S}

UV-irradiation of H$_2$S ice samples produces HS$\cdot$ radicals very readily, according to Reaction \ref{Sch.HS}. HS$\cdot$ radicals can also react with hydrogen atoms in the backwards reaction to reform the original H$_2$S molecule. IR spectroscopy shows that H$_2$S is dissociated along the first steps of UV irradiation, when recombination to form H$_2$S is inhibited, as a consequence of the low abundance of HS$\cdot$ radicals. For larger irradiation periods, however, H$_2$S recombination takes place, reducing the overall H$_2$S dissociation.\\

\begin{eqnarray}
\label{Sch.HS}
  \rm{H}_2\rm{S} \thinspace \rightleftharpoons \thinspace \rm{HS\cdot} \thinspace + \thinspace \rm{H\cdot}
\end{eqnarray}\\

The rapid formation of HS$\cdot$ radicals favors Reaction \ref{Sch.H2S2} to produce H$_2$S$_2$ molecules as well as a second hydrogen elimination (Reaction \ref{Sch.S}) forming S atoms. H$_2$S$_2$ molecules were detected at 8 K by IR spectroscopy through its H-S stretching mode centered at 2485 cm$^{-1}$ \citep{Isoniemi1999}. On the other hand, S atoms can only be detected during the warm-up phase (see Sect. \ref{sect_warm-up}).\\

\begin{eqnarray}
\label{Sch.H2S2}
  2 \thinspace \rm{HS\cdot} \thinspace \xrightarrow{} \thinspace \rm{H_2S_2}\\
\label{Sch.S}
    \rm{HS\cdot} \thinspace \xrightarrow{} \thinspace \rm{S} \thinspace + \thinspace \rm{H\cdot}
\end{eqnarray}\\

The rapid formation of H$_2$S$_2$ enables further reactions. H$_2$S$_2$ molecules can be dissociated by UV-photons according to Reaction \ref{Sch.H2Srad}. H$_2$S$_2$ molecules can also recombine with HS$\cdot$ radicals to produce H$_2$S$_3$ molecules (Reaction \ref{Sch.H2S3}).\\

\begin{eqnarray}
\label{Sch.H2Srad}
  \rm{H_2S_2} \thinspace \xrightarrow{} \thinspace \rm{H_2S\cdot} \thinspace + \thinspace \rm{H\cdot}\\
\label{Sch.H2S3}
    \rm{H_2S_2} \thinspace + \thinspace \rm{HS\cdot} \thinspace \xrightarrow{} \thinspace \rm{H_2S_3}
\end{eqnarray}\\

Finally, H$_2$S$_4$ was also detected in our experiments. The presence of HS$_3\cdot$ radicals will produce H$_2$S$_4$ by reaction with the abundant HS$\cdot$ radical (Reaction \ref{Sch.H2S4_a}). Secondly, 2 HS$_2\cdot$ radicals can react producing H$_2$S$_4$ molecules (Reaction \ref{Sch.H2S4_b}). The relatively low abundance of HS$_3\cdot$ radicals, and the low probability of 2 HS$_2\cdot$ radical-radical reactions in an environment dominated by HS$\cdot$ radicals determine the low formation rate of H$_2$S$_4$ molecules. Nevertheless, H$_2$S$_4$ was detected during warm-up of the irradiated H$_2$S ice by quadrupole mass spectrometry (QMS), as explained in Sect. \ref{sect_warm-up}.\\

\begin{eqnarray}
\label{Sch.H2S4_a}
    \rm{HS\cdot} \thinspace + \thinspace \rm{HS_3\cdot} \thinspace \xrightarrow{} \thinspace \rm{H_2S_4}\\
\label{Sch.H2S4_b}
    2 \thinspace \rm{HS_2\cdot} \thinspace \xrightarrow{} \thinspace \rm{H_2S_4} \thinspace
\end{eqnarray}\\

Non-hydrogenated molecules can also be formed. Reaction between S atoms produces S$_2$ molecules. Further S atom additions, as well as dehydrogenation of H$_2$S$_x$ species can lead to larger S$_x$ molecules. Some S$_x$ species do not absorb in the IR and others display only very weak absorption bands. These products could not be detected by IR spectroscopy in our experiments, but they were measured by QMS during warm-up.\\

\subsection{Warm-up of UV-irradiated H$_2$S ice}
\label{sect_warm-up}

Some of the radical species made during UV irradiation of H$_2$S ice react at low temperatures, such as HS$\cdot$ radicals, while larger radical species may remain in the ice matrix until thermal energy during warm-up enables subsequent reactions. IR spectroscopy confirmed the production of H$_2$S$_2$ molecules at 8 K \citep{Isoniemi1999}. Overlapping of H-S IR stretching modes of H$_2$S$_2$ with those of larger H$_2$S$_x$ species, expected to present lower formation rates, hindered the estimation of their formation temperatures.\\

\begin{figure}
  \centering 
  \includegraphics[width=0.5\textwidth]{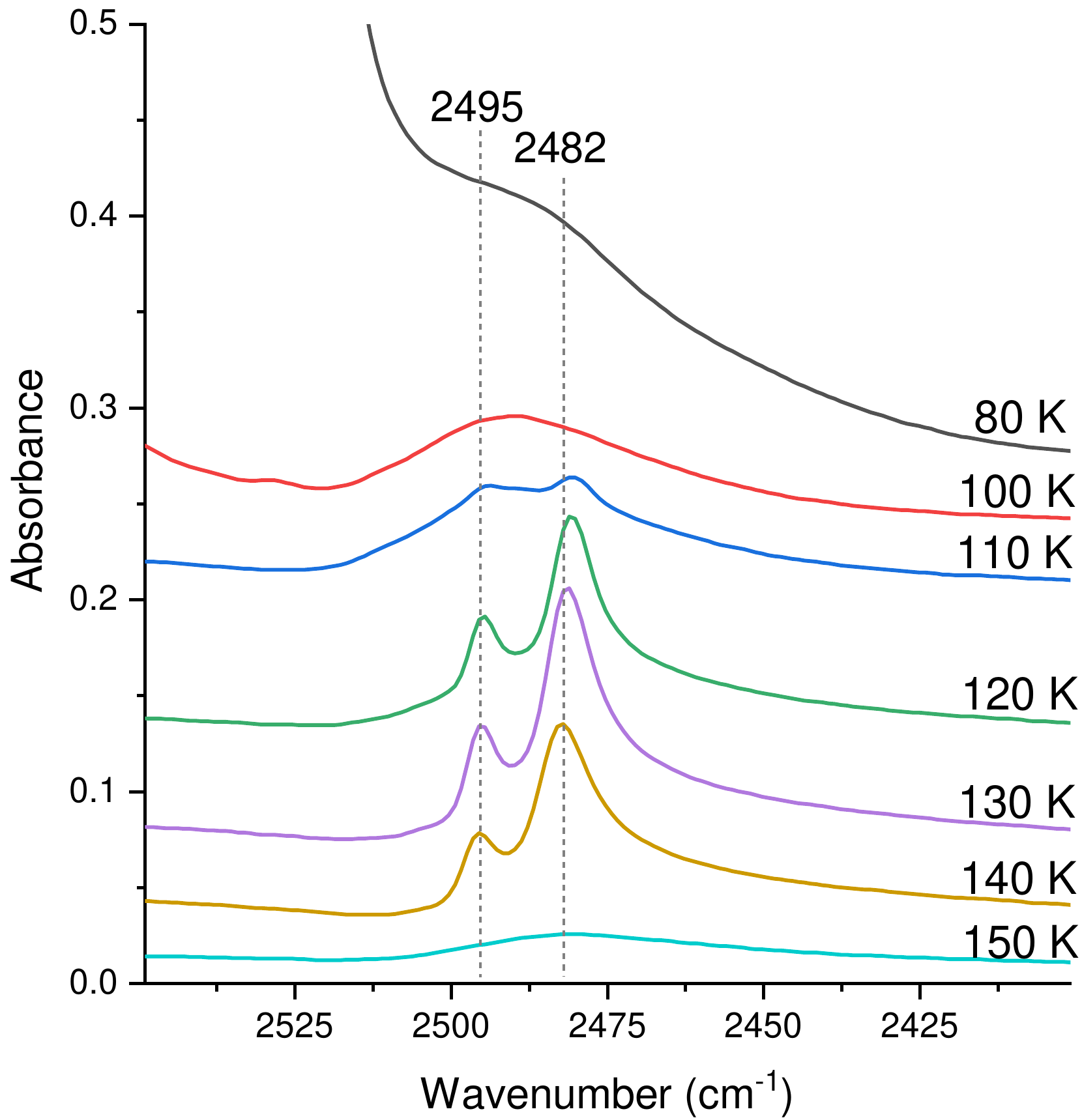}
  \caption{IR spectra collected during warm-up of a previously UV-irradiated H$_2$S ice sample. After sublimation near 90 K of most H$_2$S molecules, H$_2$S$_2$ molecules are allowed to interact forming a crystalline structure. Crystallization leads to  narrowing of the antisymmetric and symmetric vibrations of H$_2$S$_2$ that appear at 2495 and 2482 cm$^{-1}$, respectively.}
  \label{Fig.crist-h2s2}
\end{figure}

After the sublimation of amorphous and crystalline H$_2$S molecules at 96 K, IR spectra show the crystallization of H$_2$S$_2$ (Fig. \ref{Fig.crist-h2s2}). From the IR band centered around 2490 cm$^{-1}$, two absorption features can be clearly distinguished at temperatures between 110 K and 150 K. These bands belong to the antisymmetric (2495 cm$^{-1}$) and the symmetric (2482 cm$^{-1}$) vibration modes of H$_2$S$_2$. \cite{zengin1959} measured the IR spectrum of crystalline H$_2$S$_2$, reporting values in close agreement with our experiments and the work reported by \cite{Moore2007Icar..189..409M}.\\ 

The band strength of the H$_2$S$_2$ IR band at 2490 cm$^{-1}$ was estimated as follows. \cite{MD2015} provided a method to calculate the column density of a given species from the integrated QMS signal, after calibrating the QMS by using CO ice, by applying Eq. \ref{Eq.QMS}, where $N(mol)$ is the column density of a given species in molec $\cdot$ cm$^{-2}$, $A\left(\frac{m}{z}\right)$ is the integrated area taken from the QMS, $k_{CO}$ is the proportionality constant from the calibration of the QMS in a CO ice irradiation experiment leading to photodesorption of the CO molecules, $\sigma^{+}$(mol) is the ionization cross-section of each species ionized at a voltage of 70 eV in the QMS (data adopted from the National Institute of Standards and Technology), $IF(z)$ is the ionization factor, which has been considered unity for all molecules, $FF$ is the fragmentation factor, derived from the QMS spectrum of each species, and $S\left(\frac{m}{z}\right)$ is the sensitivity of the QMS \citep{MD2015}.\\ 

\begin{equation}
  N(mol) = \frac{A\left(\frac{m}{z}\right)}{k_{CO}} \cdot \frac{\sigma^{+}(CO)}{\sigma^{+}(mol)} \cdot
        \frac{IF(CO^{+})}{IF(z)} \cdot \frac{FF(28)}{FF(m)} \cdot \frac{S(28)}{S(\frac{m}{z})} \\
  \label{Eq.QMS}
\end{equation}

Then, the ratio between column density of H$_2$S and H$_2$S$_2$ (R$_\frac{H_2S}{H_2S_2}$) can be obtained from the QMS during thermal desorption of H$_2$S and H$_2$S$_2$, using Eq. \ref{Eq.QMS_2}, and taking the area below the thermal desorption peak in the QMS:

\begin{equation}
R_{\frac{H_2S}{H_2S_2}} = \frac{N(H_2S)}{N(H_2S_2)} = \frac{A(34)}{A(66)} \cdot \frac{\sigma^{+}(H_2S_2)}{\sigma^{+}(H_2S)} \cdot
        \frac{IF(H_2S_2^{+})}{IF(H_2S^{+})} \cdot \frac{FF(66)}{FF(34)} \cdot \frac{S(66)}{S(34)} \\
  \label{Eq.QMS_2}
\end{equation}\\

The band strength of the 2490 cm$^{-1}$ IR feature of H$_2$S$_2$ can be finally obtained assuming that $\sigma^{+}(H_2S_2)$ = $\sigma^{+}(H_2S)$, as there is no reported value for $\sigma^{+}(H_2S_2)$, and calculating the column density of H$_2$S from IR data before its thermal desorption:
\begin{equation}
N(H_2S_2) = \frac{1}{R_{\frac{H_2S}{H_2S_2}}} \cdot N(H_2S)
\end{equation}

Because the factor $k_{CO}$ cancels out and does not appear in Eq. \ref{Eq.QMS_2}, this method reduces the error due to i) variation in the QMS sensitivity between the CO photodesorption experiment and the H$_2$S irradiation experiment due to degradation of the QMS filament with time and ii) different environmental conditions between those experiments, since the measurement of the reference species, H$_2$S, and that of the target molecule, H$_2$S$_2$, were made during the same experiment. Entering the estimated values of $N$(H$_2$S$_2$) and its corresponding integrated IR absorption in Eq. \ref{N}, an IR band strength value of $A$(H$_2$S$_2$) = 2.4$\pm$0.5 $\times$ 10$^{-17}$ cm molecule$^{-1}$ was obtained.\\

\begin{figure}
  \centering 
  \includegraphics[width=0.5\textwidth]{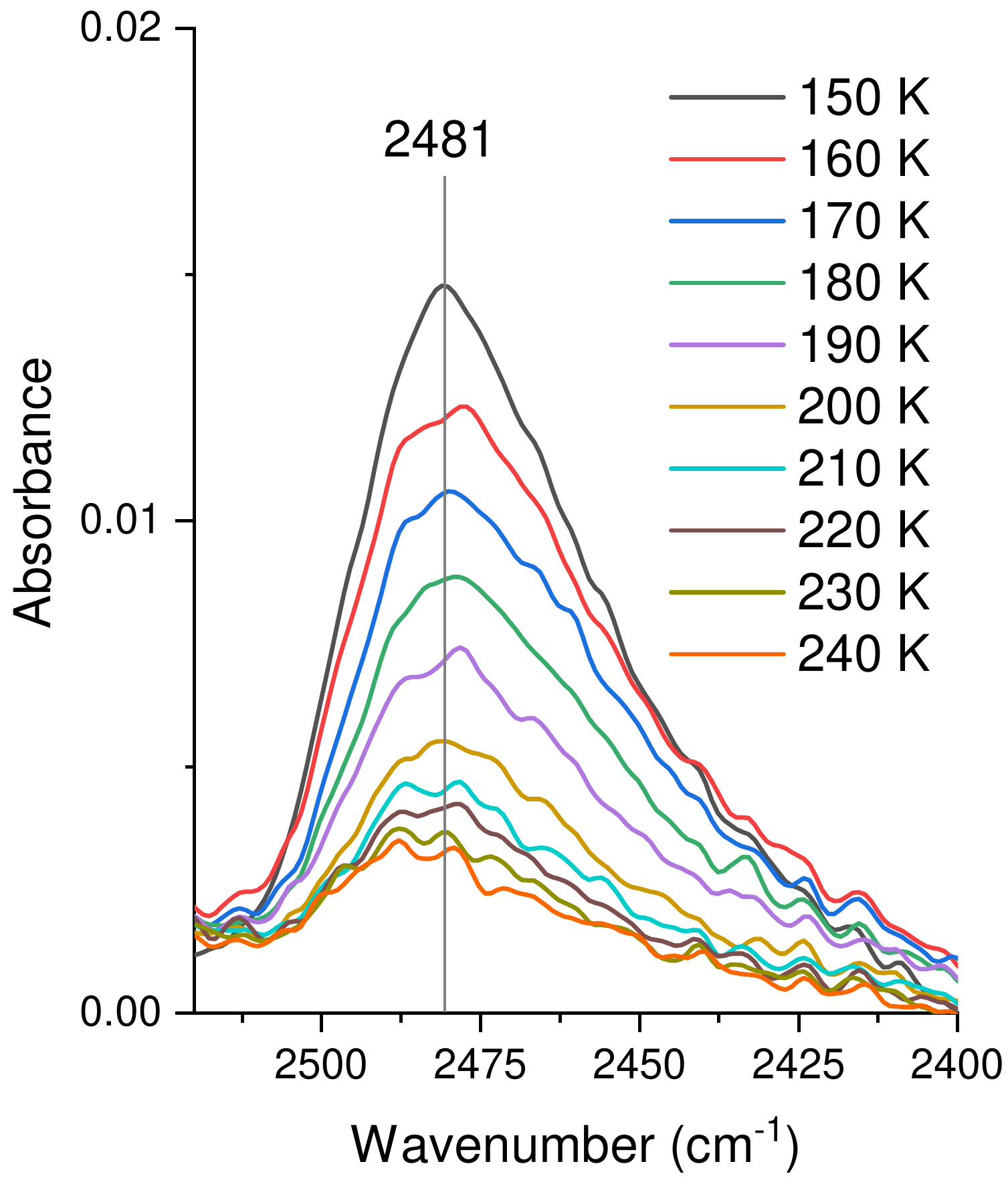}
  \caption{IR spectra acquired during warm-up of previously irradiated H$_2$S ice. \rm{This band is evidence for the presence of H$_2$S$_x$ molecules with $x$ > 2 that were detected during desorption in the gas phase by QMS, H$_2$S$_3$ and H$_2$S$_4$ desorbing at temperatures below 225 K, see Fig. \ref{Fig.QMS_H2Sx}. Therefore, the small IR absorption that remains above 240 K is likely due to H$_2$S$_x$ species with $x$ > 4 that are refractory at this temperature and could thus not be detected by the QMS.}\rm }
  \label{Fig.IR_H2Sx}
\end{figure}

After thermal desorption of H$_2$S$_2$, a new IR feature becomes visible at 2481 cm$^{-1}$ (Fig. \ref{Fig.IR_H2Sx}). This IR band disappears gradually above 150 K during the TPD, suggesting that H-S stretching modes from species less volatile than H$_2$S$_2$ contribute to this vibration mode.\\

\begin{figure}
  \centering 
  \includegraphics[width=0.5\textwidth]{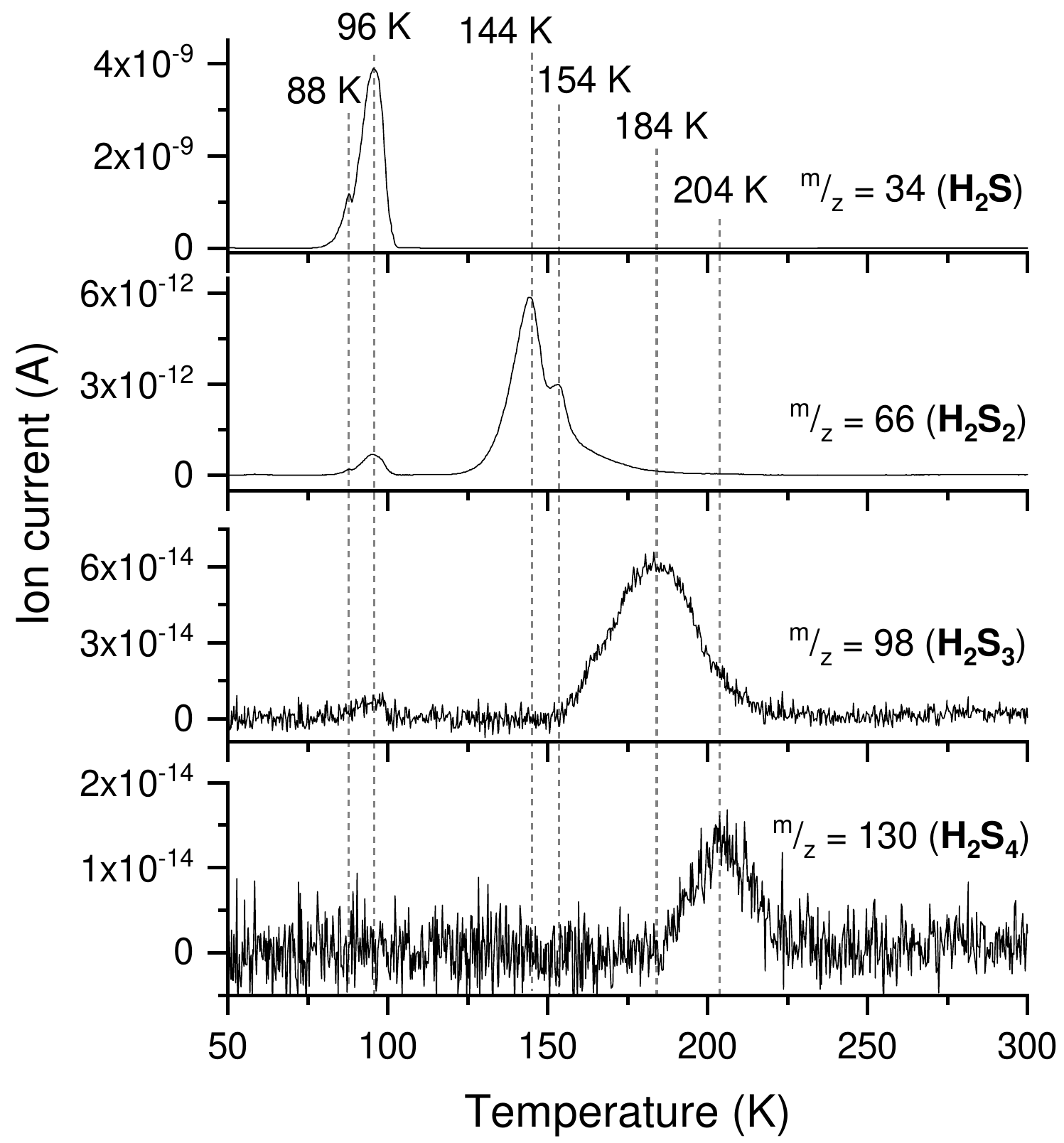}
  \caption{Thermal programmed desorption (TPD) of an irradiated H$_2$S ice sample. Up to 4 different H$_2$S$_x$ species were detected with decreasing abundances. Most intense $\frac{m}{z}$ fragment for all the species is shown.}
  \label{Fig.QMS_H2Sx}
\end{figure}

Fig. \ref{Fig.QMS_H2Sx} shows the thermal desorption of H$_2$S$_x$ molecules. Two peaks are clearly observed during H$_2$S thermal desorption. A first desorption peak appearing at 88 K coincides with the disappearance of the 2538 cm$^{-1}$ IR feature between 80 K and 90 K in Fig. \ref{Fig.IR_band_strength_H2S}, that is, thermal desorption of amorphous H$_2$S molecules. More stable crystalline H$_2$S molecules sublimate at slightly larger temperatures, producing a second peak at 96 K, see Fig. \ref{Fig.QMS_H2Sx}.

Furthermore, QMS data in Fig. \ref{Fig.QMS_H2Sx} shows evidence for H$_2$S$_2$ desorption, $\frac{m}{z} = 66$, with a maximum at 144 K, which coincides with the disappearance of the IR features at 2495 and 2482 cm$^{-1}$ attributed to this species.\\

The better sensitivity of QMS compared to IR spectroscopy allowed the identification of larger species during ice warm-up. Fig. \ref{Fig.QMS_H2Sx} shows a maximum of $\frac{m}{z} = 98$ at 184 K related to desorbing H$_2$S$_3$ molecules, and the peak of $\frac{m}{z} = 130$ at 204 K corresponds to thermally desorbed H$_2$S$_4$. The latter identification of H$_2$S$_4$ was confirmed in several irradiation and warm-up experiments.\\

\begin{figure*}
  \centering 
  \includegraphics[width=1\textwidth]{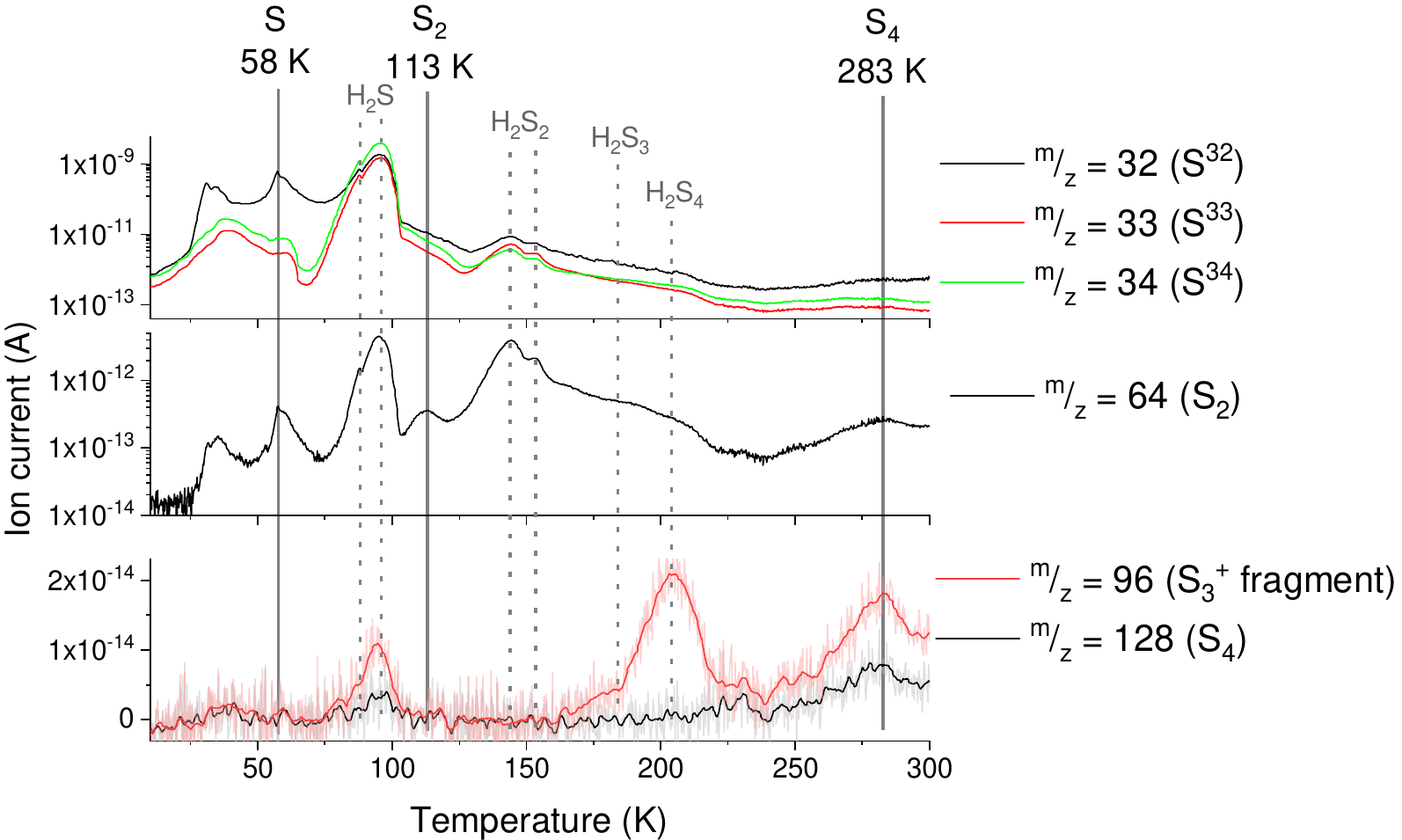}
  \caption{TPD of an irradiated H$_2$S ice sample. It should be noted that a given m/z value can correspond to the molecular ion of a species or to a fragment of a larger species. The desorption temperatures of S$_x$ molecules correspond to the solid vertical lines. Dashed lines indicate the desorption of H$_2$S$_x$ species. Apart from the molecular ion of each species, $\frac{m}{z}$ values of the main fragments used to confirm the presence of these species are displayed.}
  \label{Fig.QMS_Sx}
\end{figure*}

The fragmentation of desorbing H$_2$S$_x$ species when they impact the filament of the QMS can lead to S$_x^+$  fragments that might be erroneously attributed to the desorption of S$_x$ species formed in the ice. Nevertheless, this was not a problem since the m/z values of S$_x$ molecules were detected at temperatures that did not overlap significantly with those of H$_2$S$_x$ species. Fig. \ref{Fig.QMS_Sx} shows the thermal desorption of S$_x$ species. Thermal desorption of S atoms reaches its maximum at 58 K, dragging some S$_2$ molecules during this process. The main desorption of S$_2$ molecules takes place at 113 K. If we assume that the molecular mass is the main factor determining the desorption temperature, S$_3$ molecules are expected to desorb around 180-200 K, but they were not detected in our experiments. The non-detection of S$_3$ and the detection of S$_4$, with a maximum of desorption at 283 K, suggests that a preferential formation scheme encompass the dimerization of S$_2$ molecules, favoring the formation of S$_x$ species with a pair number of S atoms.\\

\section{Monte Carlo simulations} \label{Sect:MonteCarlo}
We used a step-by-step Monte Carlo simulation to follow the deposition of H$_2$S molecules on the surface as well as their photo-dissociation into other products. Our model is described in \cite{cazaux2015,cazaux2017}. H$_2$S molecules originating from the gas phase arrive at a random time and location on the substrate, which is represented as a grid, and follow a random walk. The arrival time depends on the rate at which gas species collide with the surface. The molecules arriving on the surface can be bound to the substrate and to other H$_2$S, HS or S species via van der Waals interactions. In the present study we use on-lattice KMC simulations. Other types of simulations, such as off-lattice KMC method allow to determine the distance of the species explicitly \citep{garrod2013}. In the present study, we consider the distance between two adsorption sites (and therefore the distance between two adjacent adsorbed species) to be equal for the entire grid, and concentrate on defining the reactions occurring within the chemical species that are deposited on the surface (H$_2$S) or are produced by photo-dissociation reactions. Our model is meant to constrain the experimental results as accurately as possible, before being exported to interstellar medium conditions in sections \ref{trans} and \ref{PS}. \\
The different mechanisms used in this model are accretion, diffusion, sublimation, chemical reactions and photo-dissociation. These different mechanisms (accretion, diffusion and sublimation) occur at rates which have been described in a previous work \citep{cazaux2018}. The accretion rate in number of molecules per second is
\begin{equation}
R_{\rm{H_2S}} = n_{\rm{H_2S}} \rm{V}_{\rm{H_2S}} \ \sigma \ \rm{Stick},
\end{equation} 
where V$_{\rm{H_2S}}= \sqrt{\frac{8\ k\ T_{\rm{gas}}}{\pi\ m_{\rm{H_2S}}}} \sim 2.5 \times 10^4 \sqrt{\frac{T_{\rm{gas}}}{100}}$ \rm{cm~s}$^{-1}$ is the thermal velocity of H$_2$S, and $\sigma$, the cross-section of the surface. Stick is the sticking coefficient that we consider to be unity in this study. The distance between two sites is assumed to be 3 \AA, which is consistent with the density of the number of sites typically assumed on surfaces N$_S$ = $({3\AA})^{-2}$  $\sim$ 10$^{15}$ cm$^{-2}$. The cross-section scales with the size of the grid considered in our calculations (which is 40$\times$40 sites) as $\sigma \sim (3 \ 10^{-8} \times 40)^2$ cm$^2$= 1.4 \ 10$^{-12}$ cm$^2$. The accretion rate of H$_2$S molecules in the experiment is 1ML/s. In our calculations, the accretion rate in ML/s can be written as R$_{\rm{H_2S}} = \frac{n_{\rm{H_2S}} \rm{V}_{\rm{H_2S}}}{NS}$ ML/s. Therefore, we chose the density of H$_2$S  in our simulations to be around $n_{\rm{H_2S}} \sim 10 ^{10}$ cm$^{-3}$ in order to reproduce experimental conditions.

\rm{The diffusion from one site to another one is described in previous work \citep{cazaux2017}. In the present study, we consider an alpha coefficient, e.g. the ratio between diffusion barriers and thermal desorption barriers, of 0.9. Such slow diffusion was determined in previous work \citep{cazaux2017} and is considered in the present study.}\rm

In the present work, we also consider photo-dissociation reactions as well as the reactions between the photo-products. These mechanisms are described in the following subsections. \\

\subsection{Photo-dissociation reactions}
Once a photon is absorbed by a H$_2$S molecule, it will dissociate it in HS + H. The product of the reaction, HS, can also receive a photon and be dissociated further in S + H. For our model, the dissociation rates are taken from \cite{Shingledecker2020} and are reported in Table \ref{diss}. As the photo-dissociation of H$_2$S can produce S atoms, these S atoms can find each other and make chains, which increases their binding energies and lowers their photo-dissociation rates as shown in Table \ref{diss}. The rates are computed as follow:
\begin{equation} 
R_{diss}=G_0*A \exp(-B\times Av) .
\end{equation}
In the experiments, each H$_2$S molecule receives a photon every 10 seconds. The parameter G$_0$ is the Habing radiation field \citep{habing1968} which is the far ultraviolet
(FUV) radiation field (where 1 G$_0$ equals a flux of 1.6 $\times$ 10$^{-3}$ erg cm$^{-2}$ s$^{-1}$, which is equivalent to 2$\times$ 10$^7$ cm$^{-2}$ s$^{-1}$). In order to have 0.1 photon per second, to mimic experimental conditions, we use G$_0$=5$\times$ 10$^6$. The dissociation rate of H$_2$S molecules is therefore R$_{diss}$= 0.015 s$^{-1}$ for H$_2$S and is of 0.006 s$^{-1}$ for HS.     

\begin{table}[htbp]
    \begin{center}
    \begin{tabular}{|c|c|c|c|c|c|}
    \hline
    React1 &  React2  & Prod1 & Prod2 & A & B\\
    \hline \hline
    HS	  & Photon	 &    H     &S  &   0.12E-08		& 2.4 \\ \hline H$_2$S  & Photon	 &    H     &HS  &   0.31E-08		& 2.6 \\ 
    \hline H$_2$S$_2$  & Photon	 &    H     &HS  &   0.55E-09		& 1.7 \\ \hline H$_2$S$_2$  & Photon	 &    H     &HS  &    0.45E-09		& 1.7 \\ \hline 
    S$_2$  & Photon	 &    S     & S  &   0.6E-09		& 1.9 \\ \hline S$_3$  & Photon	 &    S$_2$     &S  &    0.1E-09		& 0.0 \\ \hline 
    \hline S$_4$  & Photon	 &    S$_3$     &S &    0.8E-10		& 0.0 \\
    \hline S$_5$  & Photon	 &    S$_4$     &S  &    0.5E-10		& 0.0 \\ 
    \hline S$_6$  & Photon	 &    S$_5$     &S  &    0.1E-10		& 0.0\\ \hline

\end{tabular}
\caption{Photo-dissociation reactions with associated rates considered in the present study. These rates are taken from \cite{Shingledecker2020}. }
\label{diss}
\end{center}
\end{table}
 
\subsection{Chemical reactions}
 In the present simulations, we consider that species present in the solid phase (on the surface or in the ice) can react with each other. We consider a small solid phase chemical network allowing 5 possible reactions, and report the associated pre-exponential factors and barriers in table \ref{reac}. These reactions have been selected in the chemical network from \cite{Shingledecker2019}. In the present study we do not consider chemical desorption.

\begin{table}[htbp]
    \begin{center}
    \begin{tabular}{|c|c|c|c|c|c|}
    \hline
    React1 &  React2  & Prod1 & Prod2 & $\nu$ (s$^{-1}$)& E$_a$(K) \\  \hline \hline
    H	& H	    &    H$_2$     &    &  1E12		& 0 \\ \hline 
    H  & S	    &    HS        &    &  1E12		& 0 \\ \hline 
    H  & HS	    &    H$_2$S    &    &  1E12		& 0 \\ \hline 
    H  & H$_2$S	&    H$_2$     & HS &  1E12		& 1530 \\ \hline 
    H  & HS     &    H$_2$     &  S &  1E12		& 0 \\ \hline 

\end{tabular}
\caption{Reactions considered in the present study. Reactions are taken from \cite{Shingledecker2020}.}
\label{reac}
\end{center}
\end{table}

\subsection{Binding energies}
The binding energies of S atoms, as well as H, H$_2$, HS, H$_2$S and S-chains are reported in Table \ref{binding}. The binding energies for the S-chains are extrapolated by using the S-chains observed in the experiments. S$_4$ desorbs at 283~K (which implies a binding energy of $\sim$8490~K) while S$_3$ desorbs around 200~K (binding energy of $\sim$6000~K). These binding energies are reported in Table \ref{binding}. This means that the difference in desorption peak in the TPD when a S is added is around 80~K, which corresponds to a additional binding energy due to an additional S atom to the chain around ~2400~K. We take this value as an addition to the binding energy of the S-chain when an extra S atom is added. We therefore reach a binding energy of 18000~K for S$_8$. It should be noted that this is a simple assumption, and the change in binding energy could become less with the number of S atoms added, as this is the case in clusters \citep{lin2005}. 

\begin{table}[htbp]
    \begin{center}
    \begin{tabular}{|c|c|c|}
    \hline
    Species &  Binding    & Reference\\
            &  energy &          \\
            & (K) &          \\
            
            \hline \hline
    H	& 500	    &    \citealt{dulieu2005} \\ \hline 
    H$_2$  & 400	    &    \citealt{dulieu2005} \\ \hline 
    H$_2$S  & 2640	    &    \citealt{jimenez2011} \\ \hline 
    HS  & 1100	    &   \citealt{wakelam2017} \\ \hline 
    S  & 1800	    &   \citealt{wakelam2017} \\ \hline 
    HS$_2$ & 4264	    &    \citealt{jimenez2011} \\  
     & 5500	    &    \citealt{jimenez2011} \\ \hline 
    S$_2$  & 3390	 &   this work \\ \hline 
    S$_3$  & 6000     &    this work \\ \hline 
    S$_4$  & 8490     &    this work \\ \hline 
    S$_5$  & 10800     &     extrapolation \\ \hline 
    S$_6$  & 13200     &     \\ \hline 
    S$_7$  & 15600     &     \\ \hline 
    S$_8$  & 18000     &    \\ \hline 

\end{tabular}
\caption{Species considered in this study with their associated binding energies. Note that 2 values of HS$_2$ correspond to the 2 desorption peaks in the TPD implying different orientation of the molecules.}
\label{binding}
\end{center}
\end{table}

\subsection{Results}
\subsubsection{TPD simulations of H$_2$S}
\begin{figure}
   \centering
   \includegraphics[width=9cm]{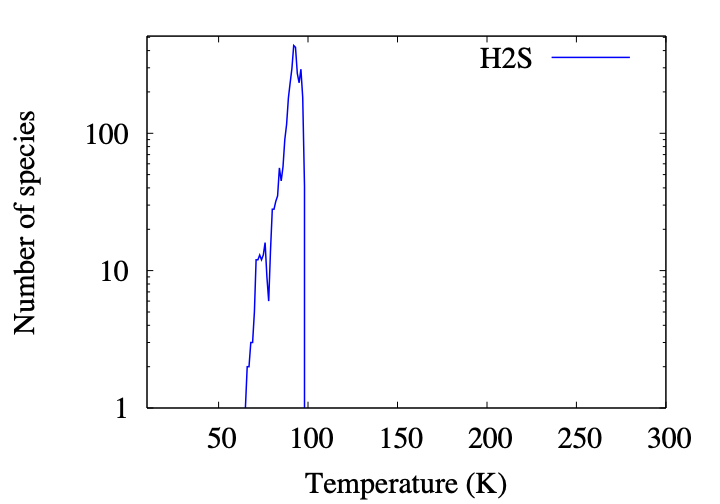}
   \caption{TPD of H$_2$S with MC simulations.}
\label{H2S}
    \end{figure}

We performed Monte Carlo simulations for the deposition and irradiation of H$_2$S in order to reproduce the experimental measurements. We first performed simulations without irradiation in order to reproduce a simple experimental TPD curve. Our results are reported in Figure \ref{H2S}. In this figure, we are able to reproduce the two peaks located around 88~K and 96~K as found in Fig. \ref{Fig.QMS_H2Sx}. To reproduce these two peaks, we considered binding energies of species in Table \ref{binding}. In our calculations, the strongest interaction between a species with a neighbor is considered. \rm{However, the width of the peaks seen in the experiments are of few tens of K. This implies that the binding energy is not a single value but a distribution of values. To reproduce the width of the peak, we consider the strongest interaction of the species (from Table \ref{binding}) and add the contribution of the neighbors around the considered species for 2\% (this value is taken as it reproduces the width the best).}\rm\ For example, an H$_2$S molecule bound with 2 neighbors would have a binding energy of 2640 + 0.02 $\times$ 2640 = 2693~K. This allows to take into account the fact that the binding energy is increased if more neighbors are involved.   

\subsubsection{TPD simulations of irradiated H$_2$S: case without ice self shielding}

\begin{figure*}
   \centering
   \includegraphics[width=6cm]{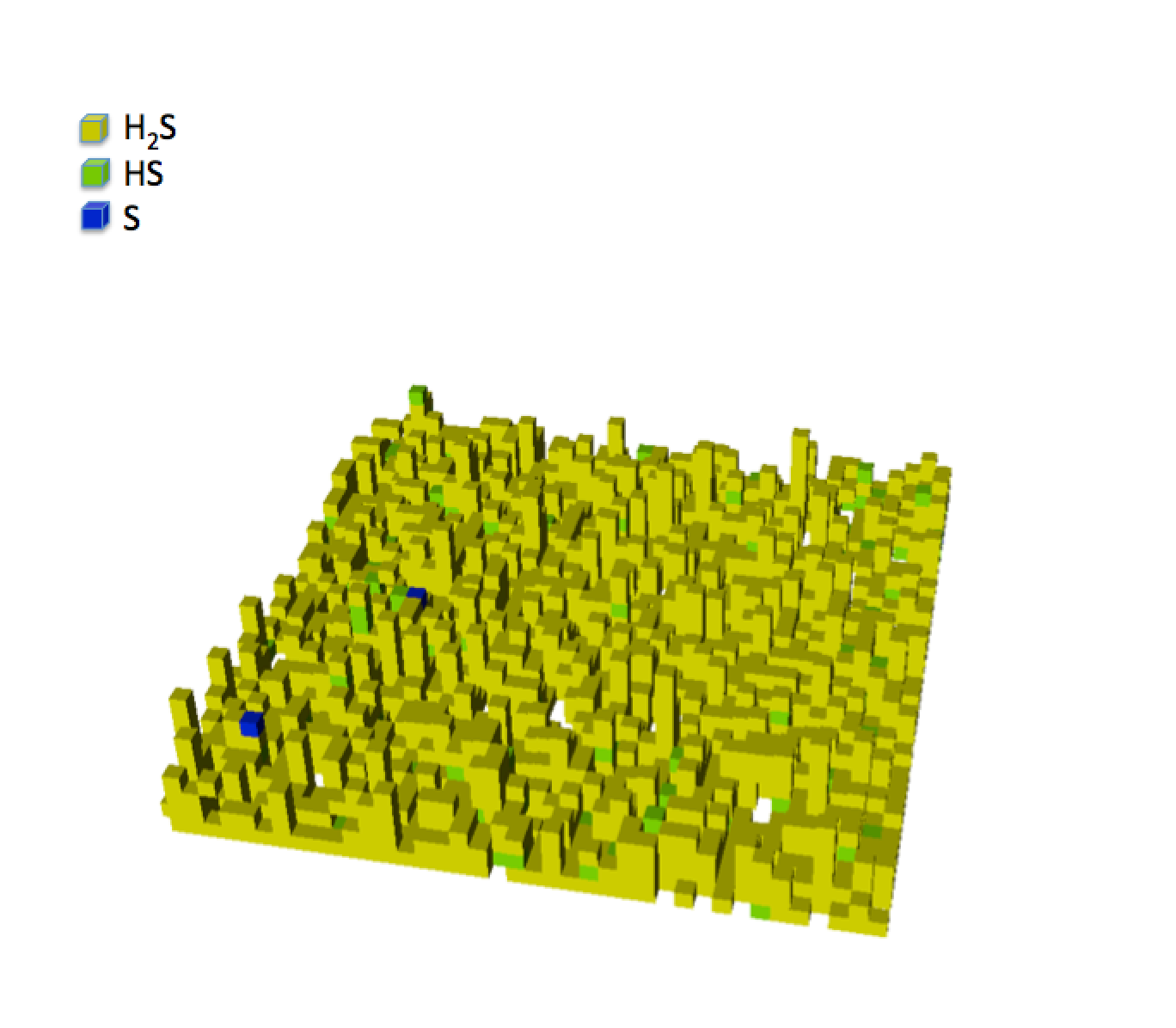}
   \includegraphics[width=6cm]{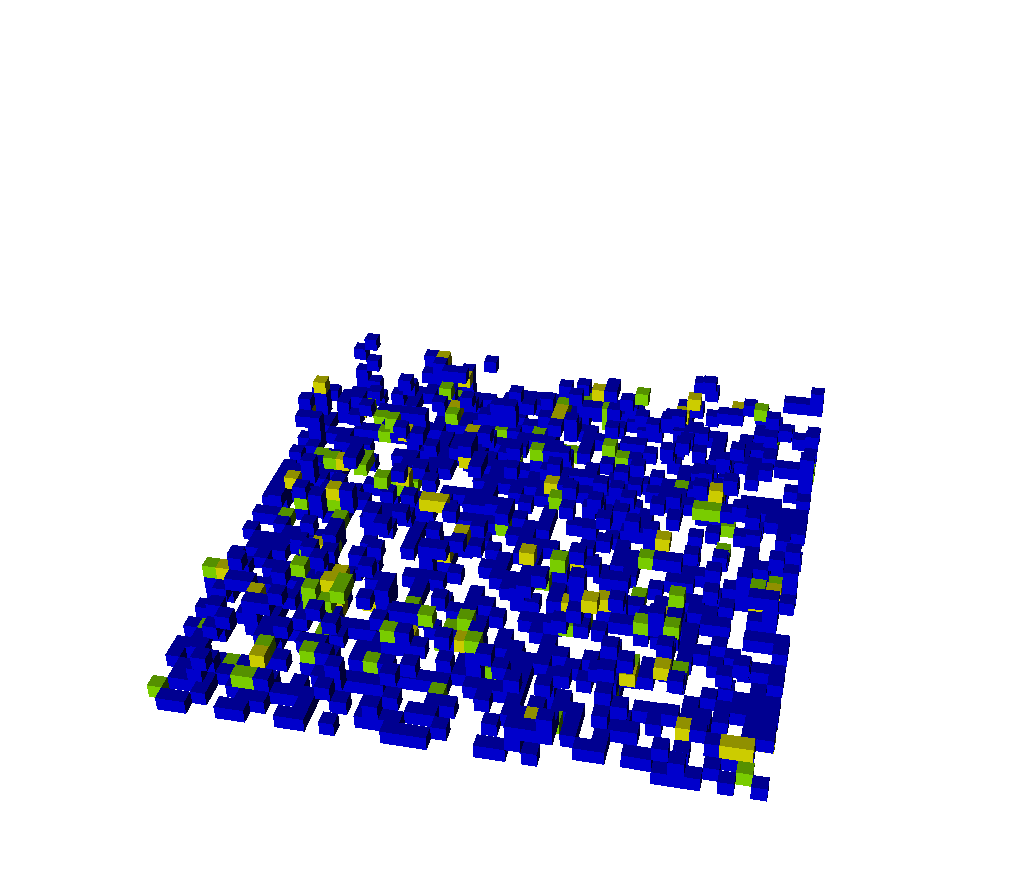}
   \includegraphics[width=6cm]{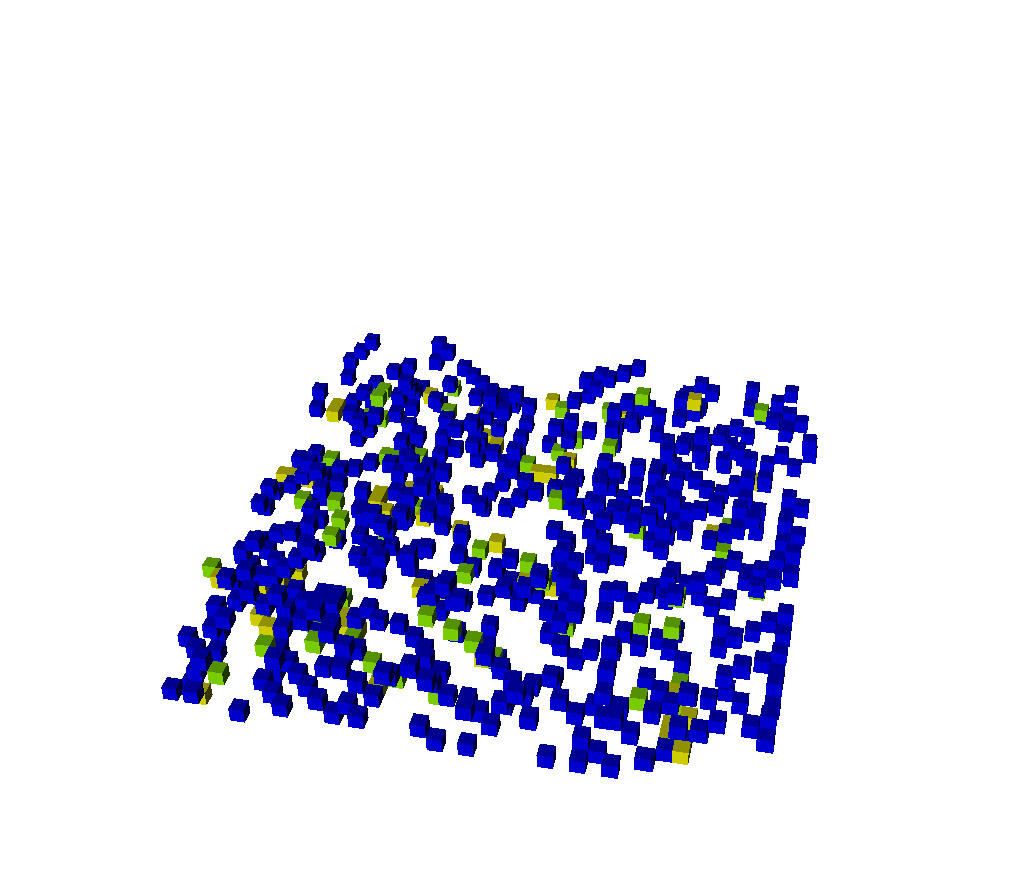}
   \includegraphics[width=6cm]{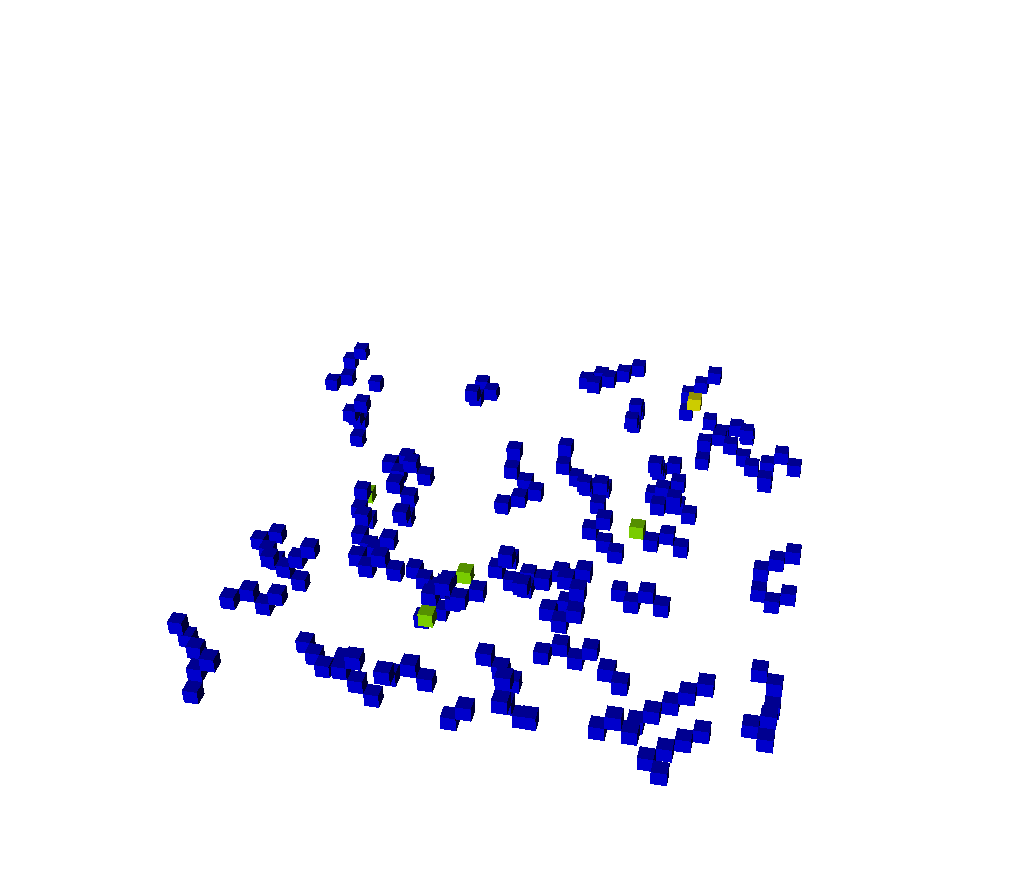}
   \includegraphics[width=6cm]{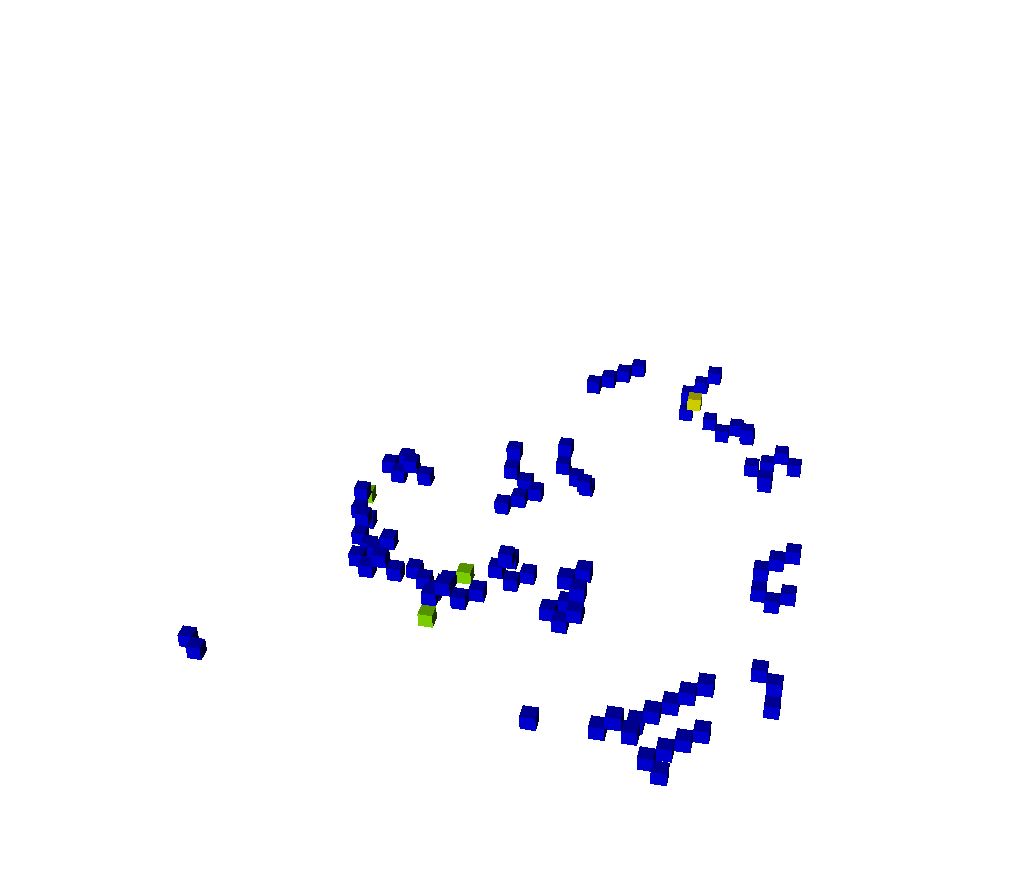}
   \includegraphics[width=6cm]{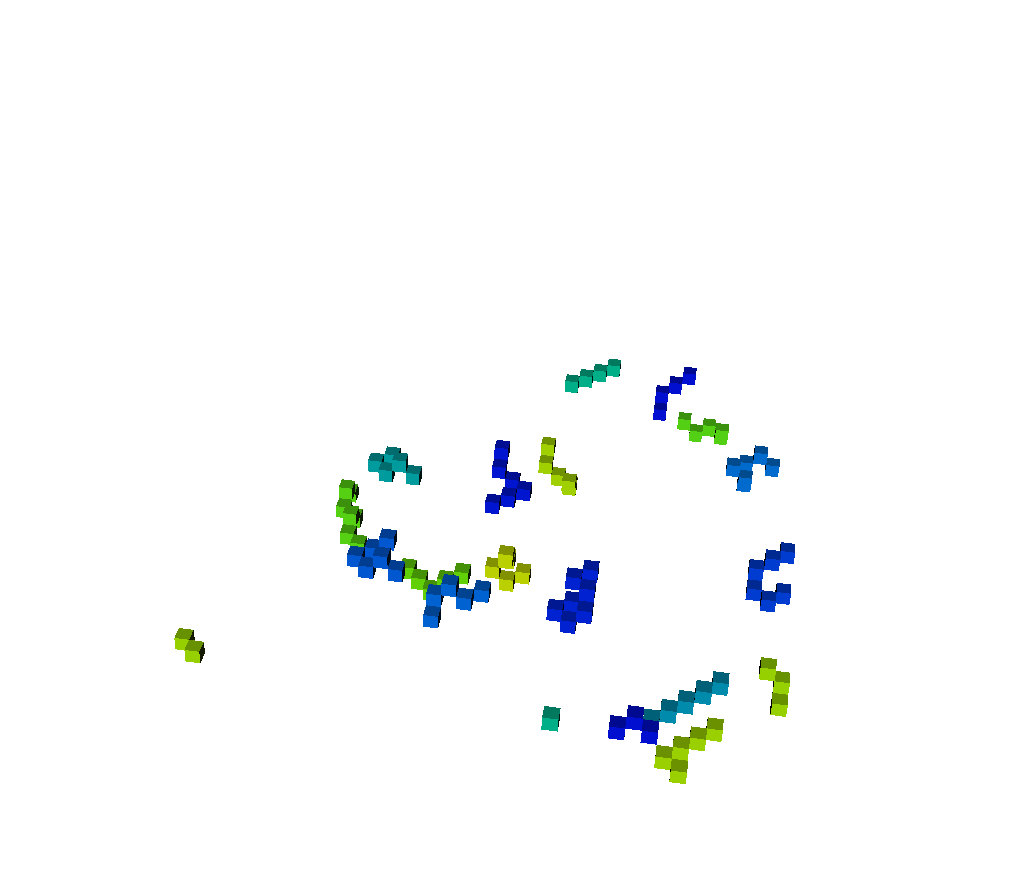}
   \caption{H$_2$S on the surface after deposition at 8~K (top left) and during TPD at 50~K (top middle), at 100~K, (top right), at 200~K (bottom left) and 300~K (bottom middle). The bottom right panel also shows the surface at 300~K but with the chains being highlighted by using one color per chain. In these simulations, the molecules in the ice do not shield each other and photons can therefore dissociate molecules from the top layers to the bottom layers. The deposition rate of H$_2$S is 1ML/s, the heating ramp is 1K/min and G$_0$=5 $\times$ 10$^6$.}
              \label{H2SUV}
   \end{figure*}

In these simulations, we deposit H$_2$S on the surface and irradiate the molecules with UV photons in rates similar to experimental rates. The vizualisation of our Monte Carlo simulations for different surface temperature is presented in Figure \ref{H2SUV}. In this figure, we show the grid of 40 by 40 sites. Each cube shows one species, the color allowing to identify the species. The H$_2$S molecules are presented in yellow, the HS in green and S atoms in blue. The first panel represents the H$_2$S after deposition, while the second, third, fourth and fifth panels show the coverage of the surface during the TPD at 50, 100, 200 and 300~K respectively. Our results show that after deposition, the H$_2$S layers have not interacted much with the photons since most of the ice is yellow (and traces of green and blue). The deposition in our simulation takes around 8 seconds, and the molecules receive a photon every 10 seconds. This means that the number of products due to photo-dissociation are less important during the deposition that during the TPD (in our simulations). This is reflected by the composition of the surface at 8~K, showing that most of H$_2$S molecules are still present on the surface with a minority of HS and S species. The surface at 50~K, shown in the top middle panel of Figure \ref{H2SUV}, illustrates a longer irradiation time, since the time spent after deposition is 42 minutes. The photo-dissociation rate of H$_2$S is R$_{diss}$ = 0.015 s$^{-1}$ in our simulations, which implies that one H$_2$S molecule would not survive such a long time since in 42 minutes it would have received around 40 photons, and the first photon would have already dissociated the molecule in HS + S. The surface coverage at 50~K is therefore mainly composed of S atoms but also HS and few H$_2$S. The photo-products HS and S are dominating compared to the coverage at 8~K after deposition. In Figure \ref{H2SUV}, top right panel, the surface temperature is of 100~K. Most of the H$_2$S and HS have desorbed from the surface, and mostly S atoms are present. The species are organized as chains, which allow them to have larger binding energies compared to individual species. These binding energies are reported in Table \ref{binding}. For high temperatures, only chains can still be present on the surface. The bottom left panel of Figure \ref{H2SUV} shows the surface at 150~K. Only chains with S atoms and sometimes an inclusion of HS or H$_2$S can be observed. At 300~K, on the bottom middle panel, only few chains are remaining on the surface and all other species have desorbed. These residues in our simulations under the form of chains cannot be seen experimentally in the TPD because their binding energies are too high. In our simulations, the longest chain has 7 S atoms, and few chains have 6 S atoms. The bottom right panel shows the same simulation and temperature than the bottom middle panel, but in this case the chains are highlighted by different colors while in the previous figures, the colors were showing which species were composing the chain (S, HS or H$_2$S). We can clearly see that some chains can reach a maximum of around 6-7 S atoms in the present conditions.

\begin{figure*}
   \centering
   \includegraphics[width=10cm]{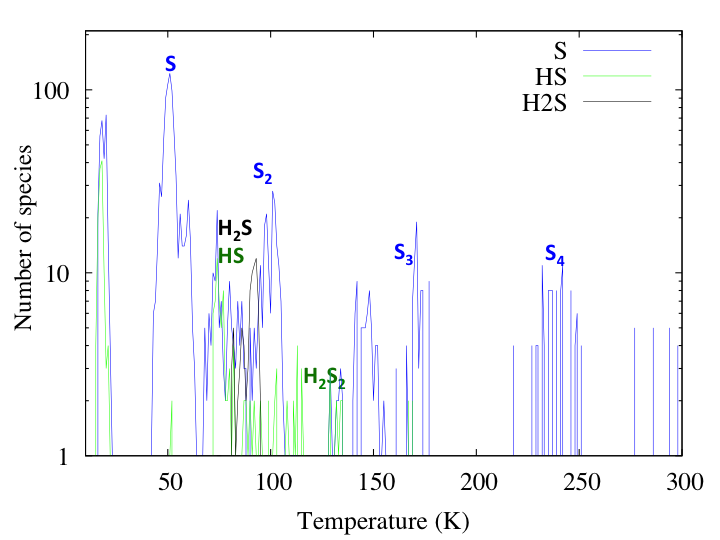}
   \caption{TPD from Monte Carlo simulations of H$_2$S. In these simulations, shielding of molecules within the ice is not considered.\rm{The deposition rate of H$_2$S is 1ML/s, the heating ramp is 1K/min and G$_0$=5$\times$ 10$^6$.}\rm}
              \label{NS_surf}
    \end{figure*}

In order to compare our results to the experimental measurements from Fig. \ref{Fig.crist-h2s2}, we present in Figure \ref{NS_surf} the simulated number of desorbed species during the TPD. Our results show different peaks that are attributed to S atoms, S chains as well as HS, H$_2$S and H$_2$S$_2$. This TPD agrees very well with the experimental results, showing chains with up to 4 members. In the present simulations, we used a coverage of few 3--5 ML (see top left panel of Figure \ref{H2SUV}). This differs from the experimental conditions where thousands of monolayers were deposited on the surface. The ratios between the different peaks of the TPD would change significantly if the simulations were run for high H$_2$S ice thickness. Our model however, cannot compute such thick ices, but in the present study we can compare the temperature of the desorption peaks in the TPD and estimate the residuals remaining in the surface at high temperatures.

  \subsubsection{TPD simulations of irradiated H$_2$S: case with self shielding}  
    
\begin{figure*}
   \centering
   \includegraphics[width=6cm]{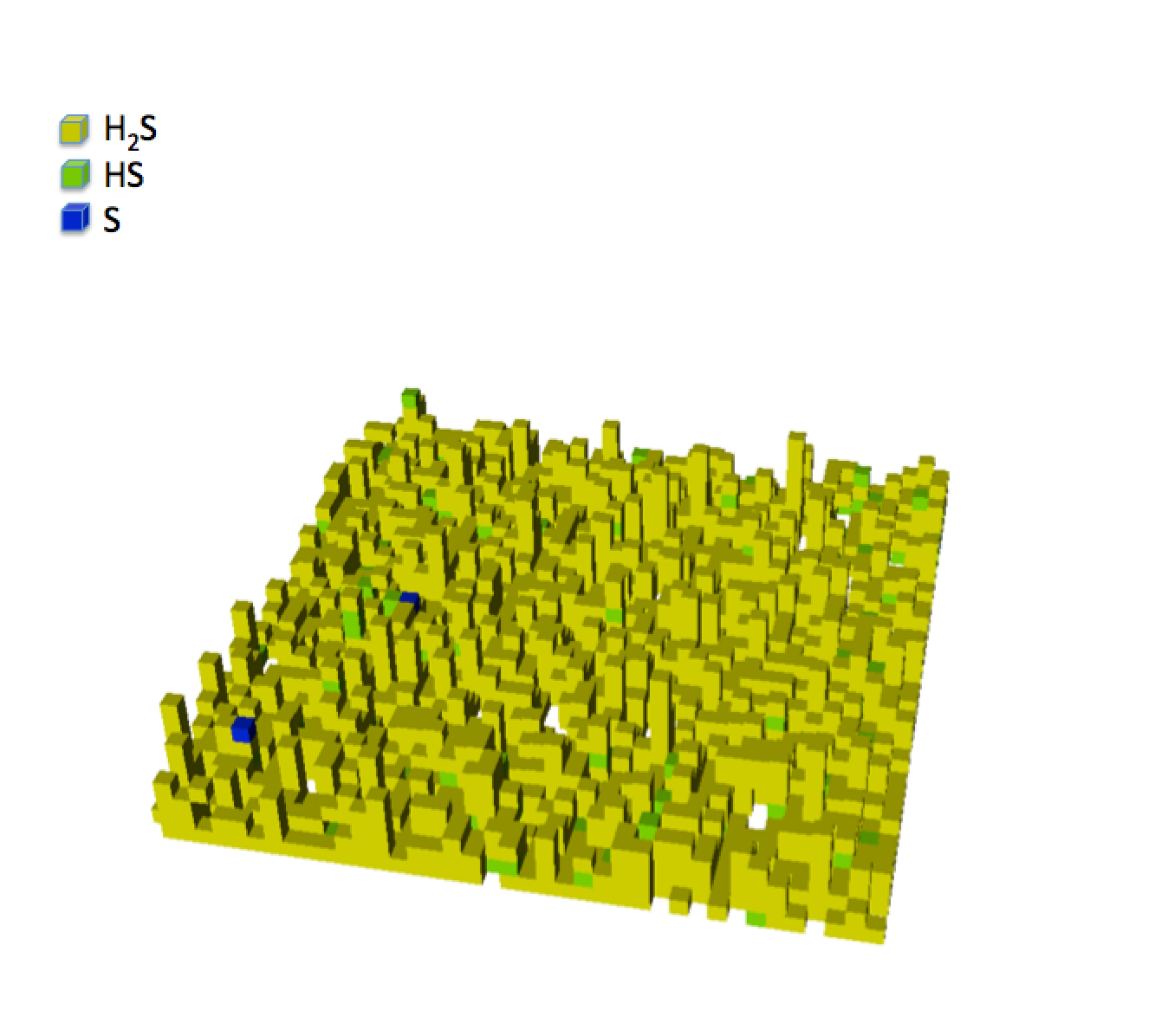}
   \includegraphics[width=6cm]{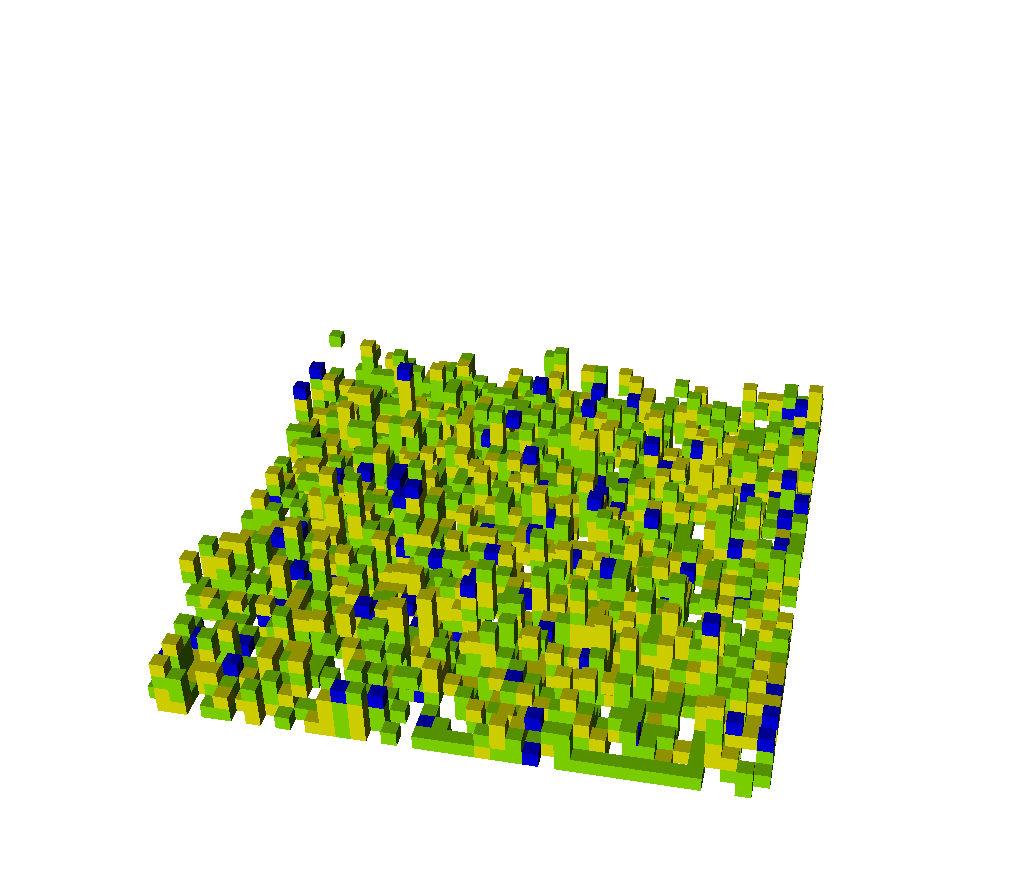}
   \includegraphics[width=6cm]{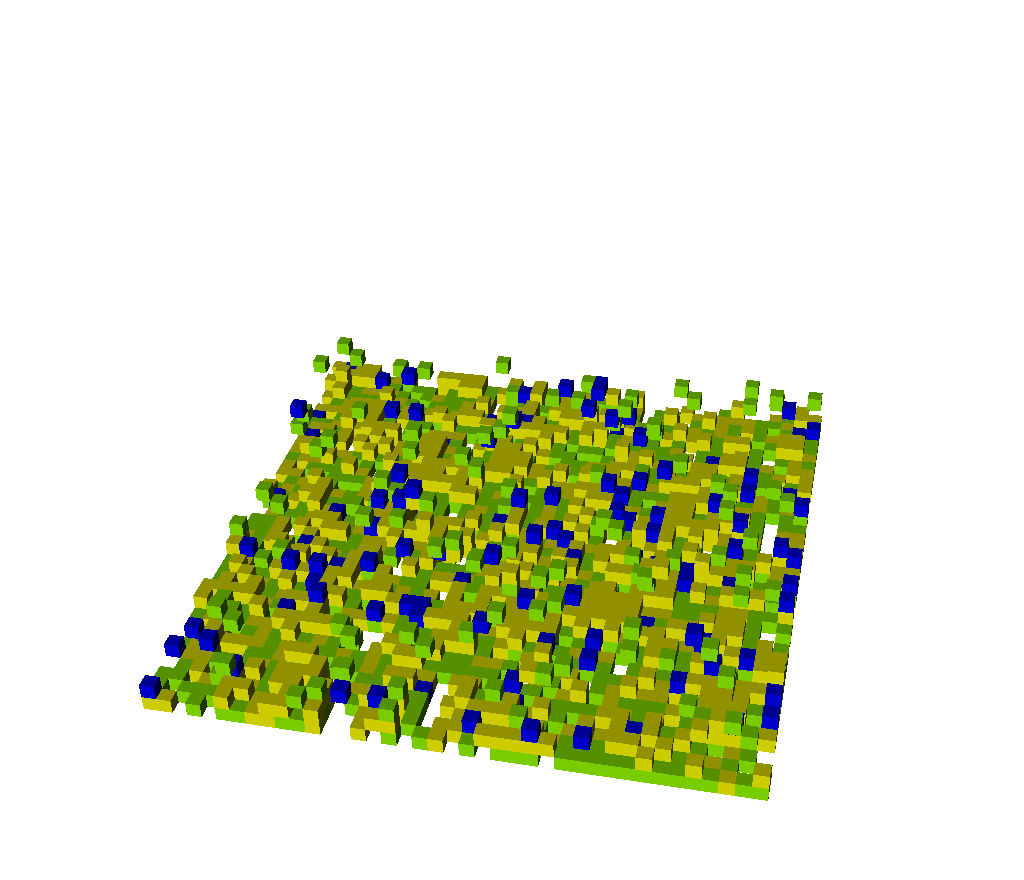}
   \includegraphics[width=6cm]{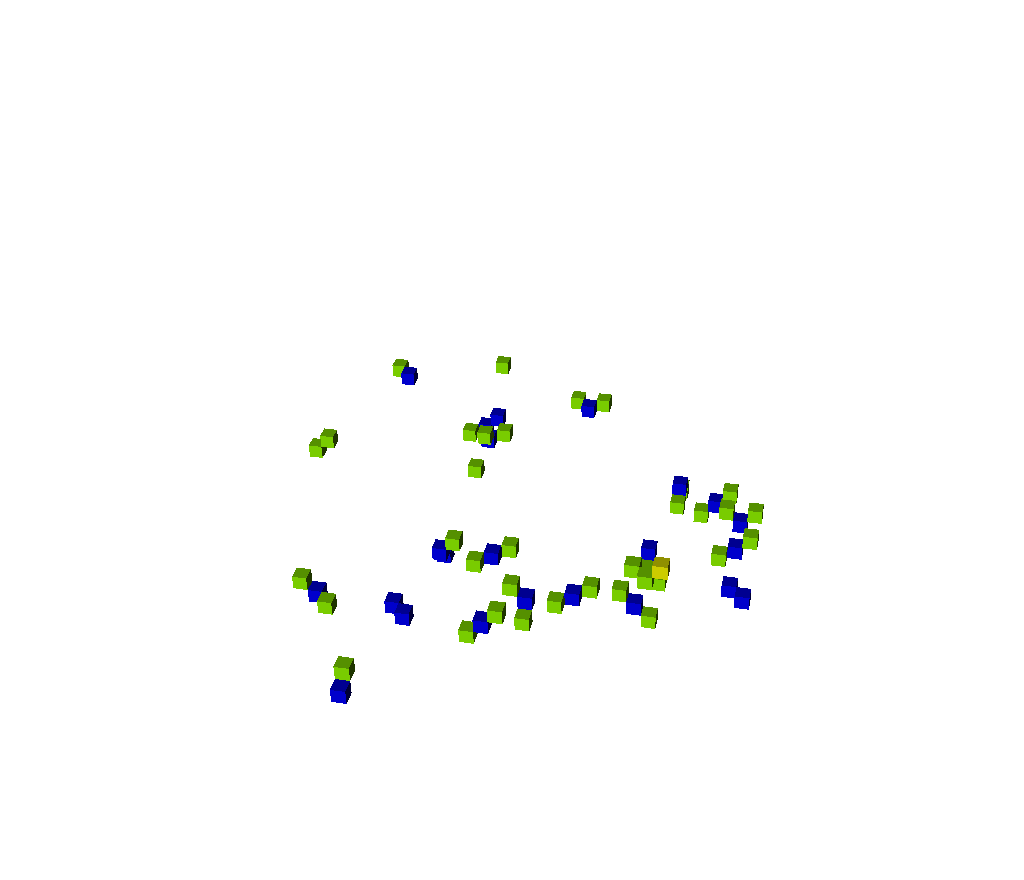}
   \includegraphics[width=6cm]{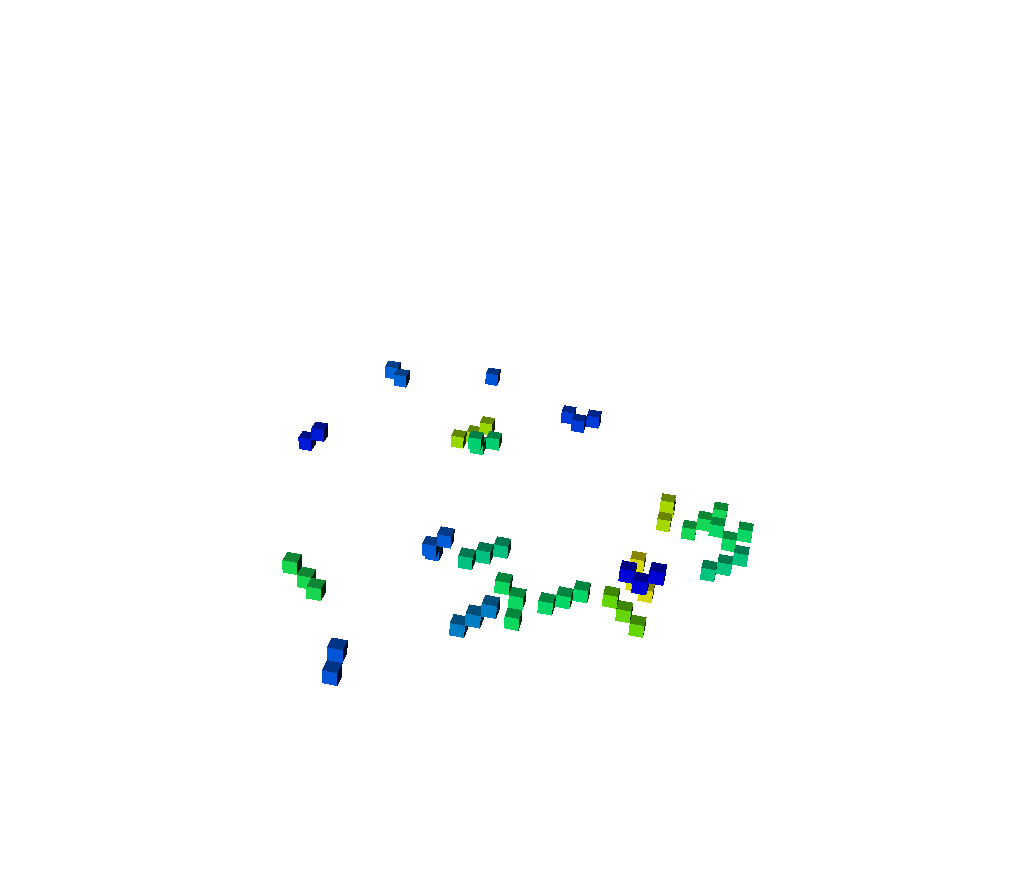}
   \caption{H$_2$S on the surface after deposition at 8~K (top left) and during TPD at 50~K (top middle), at 100~K, (top right), at 150~K (bottom left).The bottom right panel also shows the surface at 150~K but with the chains being highlighted by using one color per chain.\rm{In these simulations, the molecules in the ice do shield each other and photons cannot dissociate molecules if they are under another molecule. The deposition rate of H$_2$S is 1ML/s, the heating ramp is 1K/min and G$_0$=5$\times$ 10$^6$. }\rm}
              \label{H2SUV_S}
    \end{figure*}

\begin{figure*}
   \centering
   \includegraphics[width=10cm]{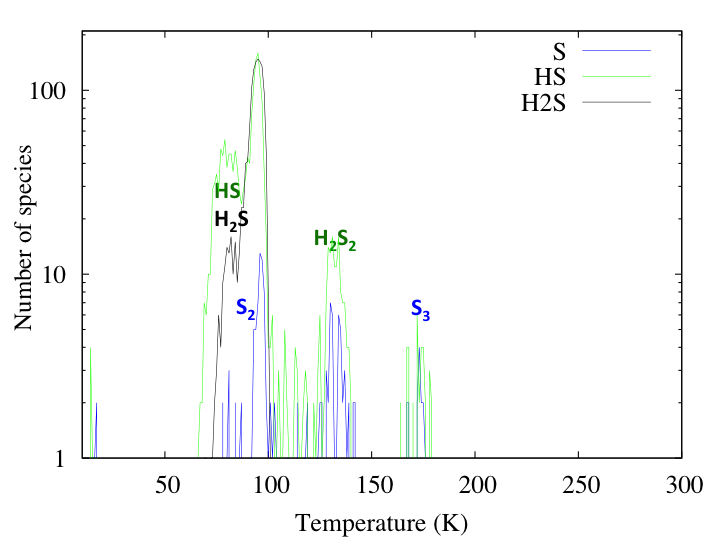}
   \caption{TPD from Monte Carlo simulations of H$_2$S. In these simulations, the molecules in the ice can shield each other. \rm{The deposition rate of H$_2$S is 1ML/s, the heating ramp is 1K/min and G$_0$=5$\times$ 10$^6$.}\rm}
              \label{S_surf}
    \end{figure*}

We perform similar simulations than in the previous section but this time, considering that molecules shield themselves from the UV radiation. For simplicity, we consider that if a species is under another one, then it is shielded from radiation and will not receive photons. It is then expected that the layers under the surface will remain H$_2$S as long as they are covered by other species, and will then be exposed only when the layers above sublimate. The results are presented in Figure \ref{H2SUV_S}. The surface coverage is shown from temperatures of 8~K (top left panel), 50~K (top middle panel), 100~K (top right panel) and 150~K (bottom left panel). These figures clearly show that there are much fewer chains on the surface than when shielded is not considered, as in Figure \ref{NS_surf}, and that instead species such as H$_2$S$_2$ (2 green boxes attached together) or H$_2$S$_3$ (1 blue box surrounded by 2 green boxes), are formed on the surface. However, no chains longer than S$_3$ are created, which is shown in Figure \ref{H2SUV_S} bottom left panel, when chains are highlighted in different colors (for a surface temperature of 150~K). Therefore, there is no species on the surface after $\sim$ 180~K as the binding energies of these chains are too low. This is reflected by the TPD in Figure \ref{S_surf}. This figure shows that S-chains are present, but only containing 2 to 3 sulfur atoms. On the other hand, many hydrogenated sulfur species are present, desorbing at temperatures lower than 180~K.       
    
\section{Photo-processes in H$_2$S ice} \label{Sect:photo}
In our simulations, we considered at first that icy molecules are not shielding each other from photons. As a result, we obtained a TPD dominated by atomic sulfur as well as S-chains, showing the presence of S-chains up to S$_4$ and a small amount of hydrogenated sulfur species. Our results show that residues are present on the surface at the end of the TPD, and that these residues are under the form of chains containing up to 6-7 sulfur atoms. Such TPD is reported in figure \ref{H2SUV}. We also considered in a second step that icy species could shield each other from UV photons, and obtained a different TPD, dominated by hydrogenated sulfur species, as shown in Figure \ref{H2SUV_S}. In this case, the species on the surface desorb below 200~K and no residue is found after the TPD. \\
The experimental results reported in Figure \ref{Fig.QMS_Sx} show that the irradiation of H$_2$S ice allows for the formation of hydrogenated sulfur species which are dominating the TPD. The S-chains are also present and are desorbing until very high temperatures. These results are reproduced by our simulations if icy species can shield each other (hydrogenated sulfur species dominate the TPD) but also if icy species do not shield each other (S-chains up to 4 S atoms in the TPD, residue on the surface after the TPD). The experimental results shows that residues are still present on the surface after the TPD. Therefore, we can assume that H$_2$S species have a small shielding properties so that hydrogenated species can survive under the present experimental conditions, and S-chains can be formed. \\
Note that we do not consider in this study the effect of photodesorption or chemical desorption. Photodesorption will reduce the amount of solid H$_2$S, meaning that less H$_2$S is available for photo-dissociation to form HS, HS$_2$ and H$_2$S$_2$ as well as chains. This implies that the total number of ice species could differ. However, the fraction of H$_2$S versus other molecules should remain similar. Chemical desorption, on the other hand, would remove some formed species and send them into the gas phase (especially via H+HS $\rightarrow$ H$_2$S \cite{Oba2018,Shingledecker2020}). This will reduce the amount of H$_2$S species compared to other chemical species in the solid phase. Our present study aims at reproducing general experimental trends in the TPDs, so neglecting these processes will only slightly affect the abundances of the species in the ices which will not affect our conclusions.

\section{Application to molecular clouds: sulfur depletion}\label{trans}

\begin{figure*}
   \centering
   \includegraphics[width=6.5cm]{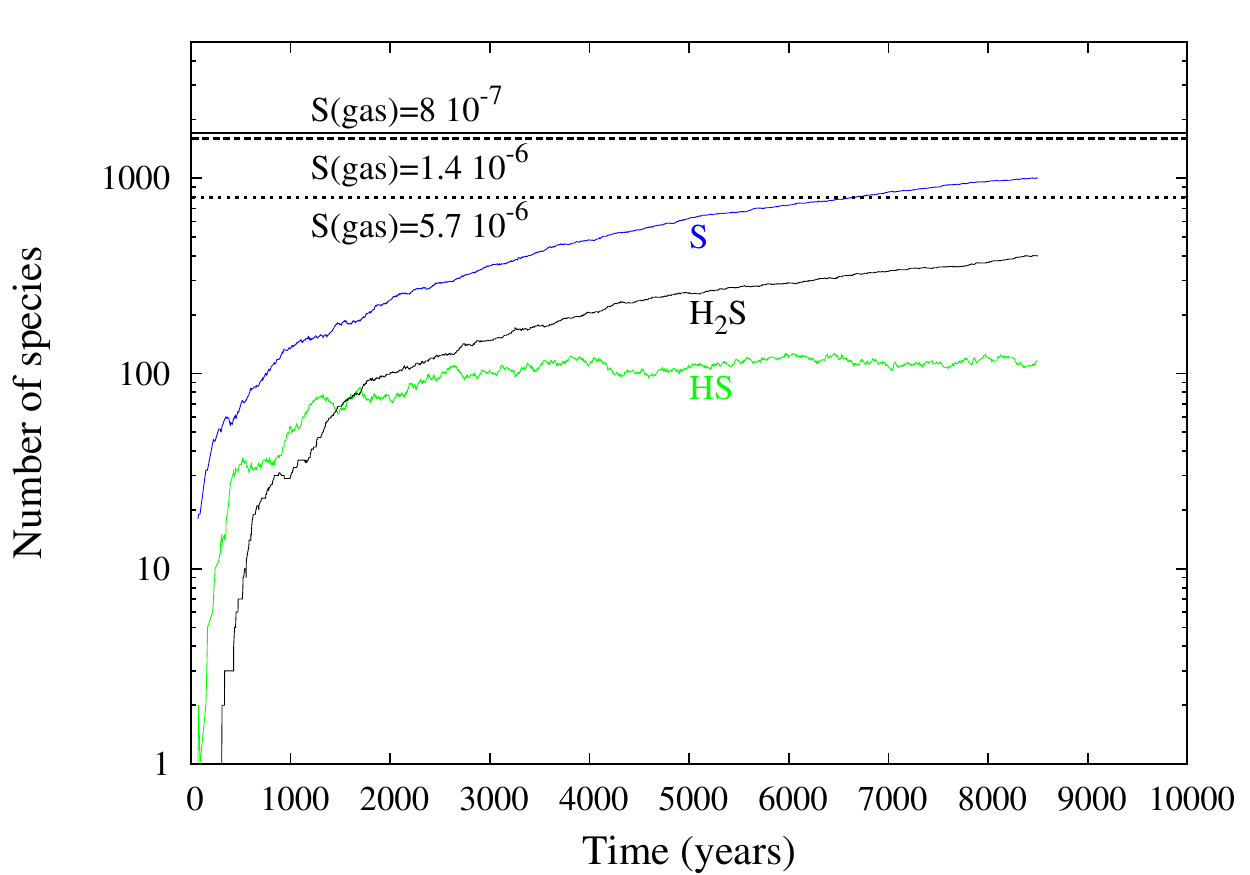}
   \includegraphics[width=6.5cm]{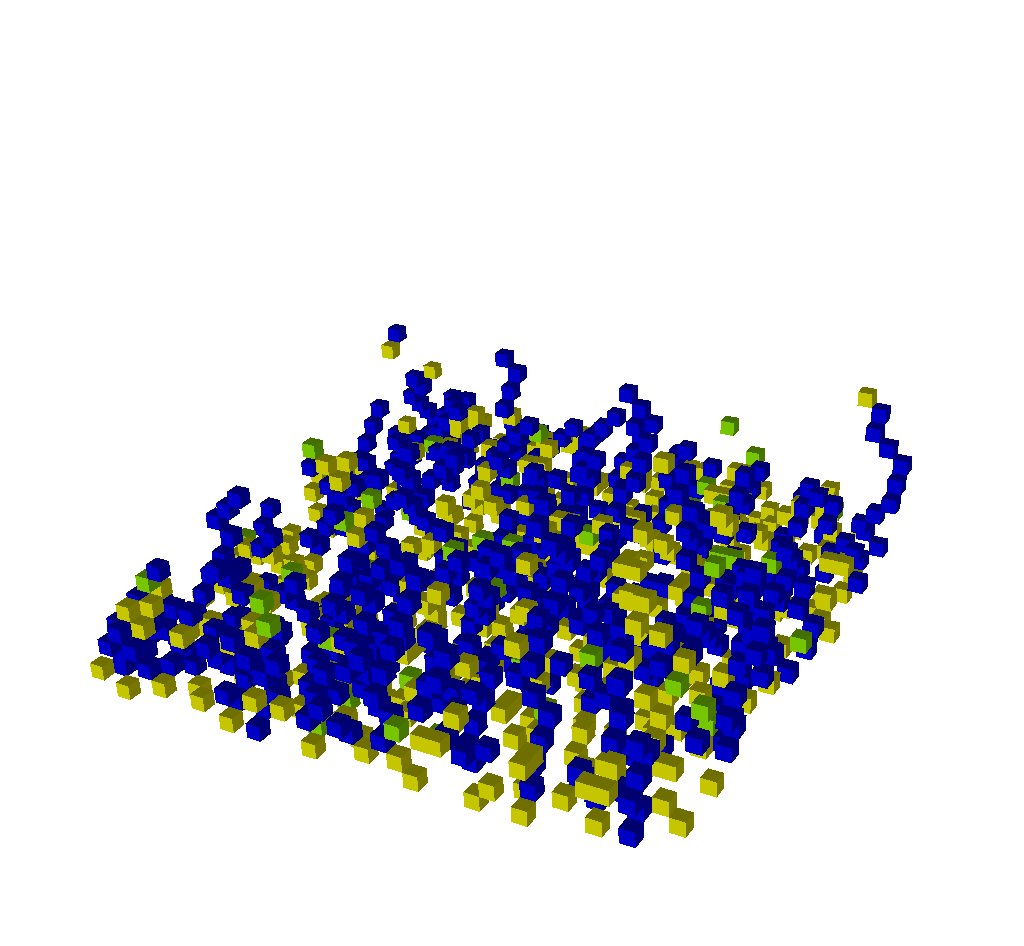}
   \caption{Sulfur coverage of S, HS or H$_2$ S as a function of time for translucent cloud models with increase of accretion rate due to the charge (left panel). The composition of the surface is visualised (right panel). \rm{In these simulations, the temperature of the dust is 17K, and G$_0$=1.}\rm}
              \label{H2SUV_S_2}
    \end{figure*}

Only $\sim$10 \% of the currently detected molecules contain sulfur atoms. This apparent lack of chemical diversity in sulfur-bearing species
 is somewhat reflective of a great problem in astrochemistry: while the observed gaseous sulfur
accounts for its total cosmic abundance (S/H$\sim$1.5$\times$10$^{-5}$) in diffuse clouds, there is an unexpected paucity of sulfur bearing molecules within molecular clouds. Indeed,
the sum of the observed gas phase abundances of S-bearing molecules (mainly SO, SO$_2$, H$_2$S, and CS) constitute only <1 \% of the 
cosmic abundance in  cold and dense cores (n(H$_2$) $>$ 10$^4$ cm$^{-3}$) (see, e.g., \citealp{Agundez2013}). One could think that
most of the sulfur is locked on the icy grain mantles in these regions but, surprisingly, the abundances of S-bearing species in the ice
as measured by near-IR observations could only account for $<$ 5\% of the total sulfur \citep{Geballe1985, Palumbo1995, Boogert1997,
jimenez2011}.  \citet{fuente2019} investigated the sulfur content
in the translucent part of TMC 1 where the freeze-out of molecules on the grain surfaces is expected to be negligible. These observations revealed that gas-phase  sulfur is already one order of magnitude lower than the cosmic value in translucent gas, i.e., 90\% of S atoms incorporate to the solid grains in the transition between the diffuse medium and the translucent cloud, corresponding to an abundance of S in the gas $\sim$8 10$^{-7}$ \citep{fuente2019}. sulfur atoms impinging in interstellar ice mantles are expected to form H$_2$S preferentially. The understanding of H$_2$S chemistry is therefore linked to the sulfur depletion problem. Chemical desorption of H$_2$S upon formation from HS + H is significant \citep{Oba2018}. \citet{navarro2020} investigated the H$_2$S abundance at the cloud edges and concluded that chemical desorption of H$_2$S from bare grains is needed to explain the high H$_2$S abundance observed in
the translucent gas  (n(H$_2$) $<$ 10$^4$ cm$^{-3}$) towards TMC 1 and Barnard 1b.\\ In the present study we concentrate on the fraction of S atoms locked onto dust particles in the translucent phase, and therefore follow the fate of S atoms impinging the dust under these conditions. We can then address whether the high depletion of sulfur could be explained by our models and more specifically, could be under the form of sulfur chains or HS or H$_2$S. In translucent clouds, sulfur is mainly under the form of S$^+$ \citep{laas2019} and in dense PDRs S$^+$ can be present until extinction of $\sim$4 (for PDRs with n$\rm{_H}$=10$^6$ cm$^{-3}$ and $\chi$=2 10$^5$, see \citealt{Sternberg1995}). The fact that sulfur is under cationic form influences its accretion rates, as discussed in \cite{Ruffle1999}. We therefore adopt a correction factor of 1+$\frac{167}{T}$ to account for the fact that the accretion rates are increased due to the charges of the sulfur as mentioned in \cite{Ruffle1999}. We performed simulations for translucent cloud conditions taking into account the fact that the accretion rate of S$^+$ is higher due to the effect of the charge. The conditions we used in our simulations are reported in Table \ref{cond} and are taken from \cite{laas2019} and \cite{snow2006}. In these simulations, reported in the left panel of Figure \ref{H2SUV_S_2}, we show the coverage of S compared to HS and H$_2$S as a function of time in translucent cloud conditions. \rm{The figure shows the increase of atomic sulfur under the form of S chains compared to H$_2$S and HS. The fact that S forms chains preferentially compared to HS or H$_2$S comes from the fact that in the gas phase, the ratio between atomic hydrogen and sulfur is 20 (see conditions in table \ref{cond}). However, due to the fact that sulfur is under cationic form, its accretion rate is increased by 1+167/T which implies an accretion rate 7.5 times faster. This means that for each S$^+$ accreting on the dust, 3 H atoms accrete. At temperatures of 17K, the residence time of H atoms is very short, and therefore S atoms can make S chains. These chains are not due to diffusion but to the fact that S$^+$ arrive on an adsorbed S atom (or a S-chain) and makes a (larger) chain.}\rm\  In order to address whether the sulfur adsorbed on the dust surface could be responsible for the depletion of sulfur in the gas phase, we estimate how much gas phase sulfur is needed to obtain the coverage obtained in our simulations. The surface area represented by carbon dust is n$_c$$\times$ $\sigma_c$ = 7.2 10$^{-22}$ cm$^{-1}$, by silicate dust is n$_s$$\times$ $\sigma_s$ = 4.4 10$^{-21}$ cm$^{-1}$, and by PAHs is $\sigma_c$ = 3.4 10$^{-21}$ cm$^{-1}$ (\citet{weingartner2001}; we consider the possibility for S-chains to form on PAHs). This corresponds to a total area of 8.5 10$^{-21}$ cm$^{-1}$. This total area is multiplied by the number of sites per area (10$^{15}$ cm$^{-2}$) to obtain the number density of sites per cm$^3$ which is 8.5 10$^{-6}$ cm$^{-3}$. This implies that if the grains would be covered by one monolayer of S (or HS or H$_2$S), then 8.5 10$^{-6}$ S atoms would be missing from the gas phase and the gas phase density of sulfur would drop to 1.4 10$^{-6}$. The abundances of sulfur in translucent clouds have been reported to be $\sim$ 8 10$^{-7}$ (this is the most likely value taking into account the uncertainties 0.4-2.2 10$^{-6}$; \citealt{fuente2019}) . Looking at our results, we show that such depletion of sulfur can occur in few times $\sim$10$^4$ years. We therefore suggest that the sulfur under the form of S$^+$ in translucent clouds plays a significant role in setting the S depletion in denser regions, as the timescales to reach such depletion are quite short. The fact that sulfur atoms are creating chains on the surface allows for layers to be built that can represent more than 100$\%$ surface coverage. The increase in S on the surface is due to the formation of S-chains. The right panel of Figure \ref{H2SUV_S_2} shows the surface coverage at the end of the simulations, the blue boxes showing the sulfur atoms and how they are organized on the surface. Our simulations suggest the creation of long S-chains under translucent cloud conditions which can explain the important depletion observed in the gas phase. As translucent clouds evolved further to molecular clouds and dense cores phases, and extinction increases, these S-rich grains will then be covered with ices. Some of these ices is composed of H$_2$S which will adsorb on the icy grains during these phases as a fraction of sulfur is still in the gas phase. The sulfur chains from the translucent cloud phase, buried underneath the ice, cannot desorb in the hot core or hot corinos phases since the temperature of the dust in these environments is too small to allow desorption (we observe up to S$_4$ until 300~K). These buried chains could be ejected again in the gas phase upon shocks that would vaporize the ice, dust and these chains. These chains could also be observed in our solar system within Interplanetary dust particles (IDPs), meteorites and comets under a refractory form.  

\begin{table}
    \begin{center}
    \begin{tabular}{|c|c|c|c|c|c|}
    \hline
    n$_H$ & n$_{HI}$/n$_H$ $^a$ & nS$^+$/n$_H$ &T$_{gas}$ (K)  & T$_{dust}$ (K)& A$_V$\\
    \hline \hline
   5 10$^3$	& 2 10$^{-4}$ & 10$^{-5}$ & 25	& 17   &    1.6  \\ \hline
\end{tabular}
\caption{Parameters used in our simulations for translucent clouds conditions. Taken from \cite{laas2019}; $^a$ taken from \cite{cazaux2009}.}
\label{cond}
\end{center}
\end{table}

\section{Application to solar system conditions: origin of sulfur found in cometary material}\label{PS}
In the present study, we show that UV irradiation of H$_2$S ice results in the formation of S chains. These chains are increasingly resistant to UV as their size increase. We show that H$_2$S species in the ice can shield other sulfur bearing species to some extent. Our study therefore shows that photo-processes of H$_2$S ices can lead to the formation of S chains that either desorb (below 250~K for chains up to S$_4$) or stay on the surface as residue for chains with more than 4 sulfur atoms.\\ Comets are known to be rich in H$_2$S ice representing $\sim$56\% of the total sulfur budget, with some minor contribution from SO$_2$, SO and S$_2$ \citep[0.14\%, ][]{calmonte2016}. The abundance of H$_2$S in comets is up to 1.5$\%$ relative to H$_2$O as inferred from millimeter and submillimeter observations
\citep{bockelee2000,Boissier2007}. In comet 67P, data analysis from Rosina (Rosetta Orbiter Spectrometer for Ion and Neutral Analysis) showed the presence of S$_3$ and S$_4$. \cite{calmonte2016} discussed whether these are real parent molecules or the products of even heavier S$_n$, fragmenting into S$_3$ and S$_4$. In the present study, we show that large S$_n$ do not desorb at temperatures below 300~K (experimental temperature range in the TPD), but stay as residue in the ice. We therefore suggest that the S$_3$ and S$_4$ chains measured in cometary material do not come from fragmentation from large S$_n$, but from UV processing of H$_2$S bearing-ices leading to small sulfur chains. 

\begin{figure*}
   \centering
   \includegraphics[width=14cm]{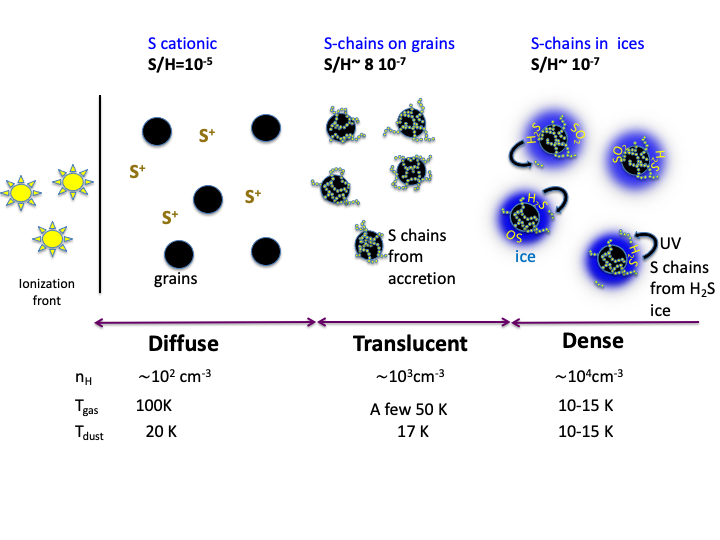}
   \caption{Sketch summarizing the results for the formation of S-chains via accretion in the translucent cloud phase, and via UV photo-processes of H$_2$S ices in molecular cloud phase. Adapted from \cite{fuente2019}.}
              \label{sketch}
    \end{figure*}

The detection of S$_2$, S$_3$ and S$_4$ makes radiolysis of H$_2$S a very likely formation process, as already discussed by other authors (\citealt{grim1987,jimeneze2012,woods2015}) which could well connect the cometary ice to the ISM. In particular, the S$_2$/H$_2$S abundance ratio in comet 67P varied from less than 1\% to about 3.4\% \citep{calmonte2016}, the latter value requires a substantial UV dose to convert H$_2$S into S$_2$ in the ice \citep{jimeneze2012}. Other species, such as H$_2$S$_2$ and S$_2$H, that are efficiently formed in H$_2$S ice  irradiation experiments \citep[][ this work]{jimeneze2012}, could not be detected in the coma of 67P; this was probably due to overlap with other species of similar mass-to-charge ratio \citep{calmonte2016}. The present study suggests that small sulfur chains come either from UV photo-processes of H$_2$S or from S depletion in the cloud prior to the formation of the protosun. \rm{The different environments at which the formation of these S-chains occur are illustrated in figure \ref{sketch}. In the translucent stage, S-chains are formed on bare grains due to the accretion of cationic S on grain surfaces, while in dense clouds, S-chains are formed via UV photo-processes of H$_2$S ice. In this case, these chains are within the ices}\rm\ Therefore, observation of small S chains alone cannot allow distinguishing between the two possible formation scenarios. Large chains, produced in the second scenario (translucent cloud phase), are present in the residues within and/or beneath ice mantles that do not desorb at temperatures typical of cometary conditions, and are therefore impossible to observe in the gas phase. Infrared bands are expected for the S$_n$ species that would be very difficult to detect since they will be masked by the intense absorption features of ice species in the same spectral region. In particular, S$_8$ displays a band near 465 cm$^{-1}$ and other bands in the far-IR \citep{trofimov2009}. These species could have been detected by the COSAC-Rosetta instrument on board the Philae lander, but the non-nominal landing on 67P comet nucleus did not allow to extract a sample for analysis (\cite{Goesmann2015}. 
Looking at the present experiments and simulations, UV photolysis of H$_2$S ices produces a fraction of H$_2$S/S$_n$ (with n $\le$3) of the order of 10$^{2}$ (from Fig. \ref{Fig.QMS_Sx}), while the S$_n$ built in the translucent cloud conditions has mainly n $>$ 4. The observations of S-chains in comet 67P seem to be related with UV processes of H$_2$S ice, rather than S-chains originating from the translucent cloud phase. This is in agreement with \cite{calmonte2016} suggesting an ice origin of the S-chains. Comparison of the different periods in the evolution of 67P show that at least in 2 periods the S follows more or less the H$_2$S abundances, suggesting a common origin in the ice matrix.

\section{Conclusions} \label{Sect:conclusions}

   \begin{enumerate}
       \item The 2538 cm$^{-1}$ band profile (H-S stretching) and integrated band strength of H$_2$S ice change gradually up to about 60 K. Above this temperature the structure is dominated by the crystalline phase, as evidenced by the distinctive narrow features of antisymmetric and symmetric S-H modes, see Fig. 1.
      \item Upon H$_2$S irradiation, the efficient formation of H$_2$S$_2$ molecules occurs already at 8 K, the lowest temperature in our experiments. Above 100 K, H$_2$S ice sublimates and H$_2$S$_2$ molecules are allowed to form a crystalline ice structure, see Fig. 2. H$_2$S$_2$ desorbs around 144 K. 
      \item Molecules H$_2$S$_x$ with $x$ > 2 are also observed in the IR, Fig. 3. This is confirmed by the desorption of H$_2$S$_3$ during warm-up near 184 K, while H$_2$S$_4$ desorbs near 204 K, see Fig. 4.
      \item During warm-up of the previously irradiated H$_2$S ice, the more volatile S$_x$ species are detected. S desorbs near 58 K, S$_2$ desorbs around 113 K, and S$_4$ desorbs at 283 K, see 
      Fig. 5. Larger S$_x$ species are more refractory and remain on the substrate at room temperature, being S$_8$ by far the most abundant species in these residues \citep{munozcaro2003}. 
      \item Monte Carlo simulations of the photo-dissociation of H$_2$S ice shows that hydrogenated sulfur species observed in the experiments could be reproduced if species in the ice can shield each other. However, this shielding should be not too efficient to be able to reproduce the S-chains observed in the experiments. We predict that the residues remaining after the TPD in the experiments are S-chains with a length up to 6-7 S atoms.
     \item Applying our Monte Carlo simulations to translucent clouds conditions, we show that long S-chains are created in a rather short timescales of few 10$^4$ years. The important depletion observed in these environments can therefore be explained by grains covered by S$_n$. These grains will be buried under ices in later stages as the cloud evolves.
    \item Small S-chains observed in comets could in theory originate from processed H$_2$S ices (from molecular cloud phases) or from S-chains from the translucent cloud phase. Since small chains are being mainly produced by photo-processes of H$_2$S ice, we suggest that observations of S$_3$ and S$_4$ in 67P have an icy origin, and are not due to processes of long chains from the translucent cloud phase. 
   \end{enumerate}

\begin{acknowledgements}
      Part of this work was supported by the German
      \emph{Deut\-sche For\-schungs\-ge\-mein\-schaft, DFG\/} project
      number Ts~17/2--1. H. C. was supported by PhD fellowship FPU-17/03172. AF, PRM and DNA thank the Spanish MICINN for support under 
      grant PID2019-106235GB-I100.
\end{acknowledgements}

%
%

\bibliographystyle{aa} 

\end{document}